\documentclass[aps,prb,twocolumn,groupedaddress,showpacs,preprintnumbers]{revtex4-1}

\usepackage{relsize}
\usepackage{mathtools}
\usepackage{amssymb}
\usepackage{amsmath}
\usepackage{braket}
\usepackage{pgf}

\newcommand{\eM}     {$\epsilon$-machine}

\newcommand{\CHC} {{ChC}}

\newcommand{\csg} {{coronal spectrogram}}
\newcommand{\Csg} {{Coronal spectrogram}}

\newcommand{\e} {{\mathrm{e}}}

\newcommand{\Hagg} {H\"{a}gg}

\newcommand{\TransMatHagg} {{\sf{T}}}
\newcommand{\TransMatABC} {{\mathcal{T}}}
\newcommand{\IdentMat} {{\mathbb{I}}}
\newcommand{\abcT} {\TransMatABC} 
\newcommand{\HaggT} {\TransMatHagg}
\newcommand{\HaggTzero} {\HaggT^{[0]}} 
\newcommand{\HaggTone} {\HaggT^{[1]}} 
\newcommand{\abcTA} {\abcT^{[A]}} 
\newcommand{\abcTB} {\abcT^{[B]}} 
\newcommand{\abcTC} {\abcT^{[C]}} 
\newcommand{\TransMatSet} {{\mathcal{\mathbf{T}}}}
\newcommand{\TransMat} {{\mathcal{T}}}

\newcommand{\symbolABC} {{x}}

\newcommand{\StateABC} {{{\mathcal S}}}

\newcommand{\Machine} {{\Gamma}}
\newcommand{\Alphabet} {{\mathcal A}} 
\newcommand{\States} {{\mathbb{S}}}
\newcommand{\DP} {{\sf{I}}}
\newcommand{\AlphabetABC} {{{\mathcal A}_{\rm P}}}

\newcommand{\MachineABC} {{$ABC$-machine}}
\newcommand{\MachineHagg} {{H\"{a}gg-machine}}
\newcommand{\Dist} {{ \boldsymbol{\pi} }}
\newcommand{\DistOne} {{\boldsymbol{1}}}
\newcommand{\One} {{\boldsymbol{1}}}

\newcommand{\Nstates} {M}
\newcommand{\NstatesHagg} {{\sf M}_{\rm H}}
\newcommand{\NModLayers} {{N}}
\newcommand{\coperator} {{\hat{{\rm {c}}}}}
\newcommand{\aoperator} {{\hat{{\rm {a}}}}}
\newcommand{\soperator} {{\hat{{\rm {s}}}}}
\newcommand{\xoperator} {{\hat{\xi}}}

\newcommand{\realpart} {{\mbox {\, \large{$\Re$}}}}

\newcommand{\ie} {{\it {i.e.}}}
\newcommand{\etal} {{\it {et~al.}}}
\newcommand{\Qxin} {{Q_{\xi}(n)}}
\newcommand{\Qcn}  {{Q_{\rm {c}}(n)}}
\newcommand{\Qan}  {{Q_{\rm {a}}(n)}}
\newcommand{\Qsn}  {{Q_{\rm {s}}(n)}}

\newcommand{\CFBraKet}{\Braket{\abcT_\lambda ^{\xoperator(\Alphabet)}}}

\newcommand{\CFBraKetm}{\Braket{\abcT_{\lambda, m} ^{\xoperator(\Alphabet)}}}

\begin{document}


\title{Diffraction Patterns of Layered Close-packed Structures\\
from Hidden Markov Models}

\author{{P.~M.~Riechers}}
\email[]{pmriechers@ucdavis.edu}
\affiliation{Complexity Sciences Center \& Physics Department,
	University of California, One Shields Avenue,
	{Davis, California} 95616, {USA}}

\author{D.~P.~Varn}
\email[]{dpv@complexmatter.org}
\homepage[]{http://wissenplatz.org/}
\affiliation{Complexity Sciences Center \& Physics Department,
	University of California, One Shields Avenue,
	{Davis, California} 95616, {USA}}

\author{J.~P.~Crutchfield}
\email[]{chaos@ucdavis.edu}
\homepage[]{http://csc.ucdavis.edu/$\sim$chaos/}
\affiliation{Complexity Sciences Center \& Physics Department,
	University of California, One Shields Avenue,
	{Davis, California} 95616, {USA}}


\date{\today}

\begin{abstract}
We recently derived analytical expressions for the pairwise (auto)correlation
functions (CFs) between modular layers (MLs) in close-packed structures (CPSs)
for the wide class of stacking processes describable as hidden Markov models
(HMMs) [Riechers \etal, (2014), Acta Crystallogr.~A, XX 000-000]. We now use
these results to calculate diffraction patterns (DPs) directly from HMMs,
discovering that the relationship between the HMMs and DPs is both simple and
fundamental in nature. We show that in the limit of large crystals, the DP is a
function of parameters that specify the HMM. We give three elementary but
important examples that demonstrate this result, deriving
expressions for the DP of CPSs stacked (i) independently, 
(ii) as infinite-Markov-order randomly faulted
2H and 3C stacking structures over the entire range of growth and
deformation faulting probabilities, and (iii) as a HMM that models
Shockley-Frank stacking faults in 6H-SiC. While applied here to planar
faulting in CPSs, extending the methods and results to planar disorder in other
layered materials is straightforward. In this way, we effectively solve the
broad problem of calculating a DP---either analytically or numerically---for
any stacking structure---ordered or disordered---where the stacking process can
be expressed as a HMM.
\end{abstract}


\preprint{Santa Fe Institute Working Paper 14-10-XXX}
\preprint{arxiv.org:1410.XXXX []}

\maketitle

\section{Introduction}
\label{Introduction}

Increasingly, materials scientists appreciate the sometimes unexpected role
that disorder and crystal defects play in material properties. Crystal defects
have, of course, been known and studied for some time,~\cite{Hirt68a} but in
the past they have often been viewed as a nuisance or feature to be minimized or
eliminated.~\cite{Caro11a} However, as material scientists acquired the
ability to engineer the composition and structure of materials at the nanoscale
level,~\cite{Geim13a} they have concomitantly discovered that disorder, far
from being unwelcome, can result in specimens with desirable material
properties.  For example, the introduction of defects into graphene nanosheets
improves their performance in batteries,~\cite{Pan09a} and so-called `defect
engineering' in semiconductors is attracting wide attention.~\cite{Seeb10a}

The growing technological import of disordered materials then challenges the
crystallographic community to develop a theoretical framework capable of
discovering, describing, and quantifying the noncrystallinity so often present
in condensed matter systems.  While concepts such as the Bravais lattice and
point and space groups have allowed for the classification and codification of
perfectly ordered materials, a new formalism---founded on concepts and
mathematical constructs that intrinsically treat nonzero entropy systems---is
needed.~\cite{Mack86a,Cart12a} For quasi-one-dimensional materials, this
formalism has been identified: \emph{chaotic crystallography}
(\CHC).~\cite{Varn14a}

\CHC\ is the use of information- and computation-theoretic methods to discover,
describe, and categorize material structure. Specifically, it adapts and
applies \emph{computational mechanics}~\cite{Crut12a,Shal01b} to the problem
of disorder in materials. Drawing from concepts developed in information
theory,~\cite{Cove06a} theoretical computer science~\cite{Paz71a,Hopc79a} and
nonlinear dynamics,~\cite{Stro01a,Feld12a} computational mechanics has been
successfully applied to a number of physical
systems.~\cite{Palm00a,Clar01a,Kell12a}

While many kinds of disorder may be present in materials, here we restrict our
attention to the disorder that results from shifting or displacing an entire
plane of atoms. We assume that the three-dimensional material is built up from
the stacking of identical, crystalline, two-dimensional layers, which here we
refer to as \emph{modular layers} (MLs).~\cite{Pric83a,Ferr08a} There are usually
constraints on the allowed alignment between adjacent MLs, effectively
restricting to a small set the number of possible ML orientations. Thus, a
complete description of a quasi-one-dimensional disordered specimen reduces to
a one-dimensional list of the successive orientations of the MLs encountered as
one moves along the stacking direction, called the \emph{stacking
sequence}.~\cite{Varn02a} The effective stochastic process induced by scanning
the stacking sequence is referred to as the \emph{stacking
process}.~\cite{Varn02a} 

For specimens containing many MLs, listing the entire stacking sequence can be
cumbersome, as well as unnecessary. Since many properties of disordered
materials depend not on the specific stacking sequence, but rather on the
statistics of the stacking sequence---\ie, the stacking process---\CHC\
conveniently specifies material structure in terms the hidden Markov model
(HMM) that describes the stacking process. We contend that any quantity
depending on the statistics of the stacking process should be amenable to
direct calculation from the process's HMM. Indeed, it has recently been
shown~\cite{Riec14b} that the pairwise (auto)correlation functions between MLs
are directly calculable from the HMM. However, until now, several quantities,
such as the diffraction pattern (DP), have defied exact calculation from a
general HMM. Of course, the difficulty of estimating DPs from models of
disordered stackings predates the use of HMMs. In fact, there is a considerable
literature devoted to deriving expressions for the DP for faulted specimens or
special cases of
disorder.~\cite{Hend42a,Wils42a,Kaki65a,Holl69a,Holl69b,Pand80a,Pand80b,Pand80c,Seba94a,Gosk01a,Kopp12a}
Finite-order Markov processes have a formal solution in the work following
Hendricks and Teller~\cite{Hend42a} and Kakinoki and Komura~\cite{Kaki65a}, but
even simple examples with two coexisting faulting mechanisms transcend the
limitations of these methods. Researchers typically work out the consequences
for particular models, but a general theory has been elusive.

To circumvent the often tedious algebra innate to these analytical techniques,
Berliner and Werner~\cite{Berl86a} demonstrated that CFs and thereby the DP
could be explicitly calculated by taking a sample stacking sequence derived
from a particular model of disorder. Though now in common usage, this approach is not altogether satisfactory. The first difficulty lies in the
statistical fluctuations inherent from considering any finite-size sample. It
is known that these statistical fluctuations lead to fluctuations in the power
spectrum that are on the order of the magnitude of the power spectrum
itself.~\cite{Kant04a} This difficulty can be ameliorated by taking many
samples and averaging~\cite{Niko89a} or by using a smoothing
procedure.~\cite{Varn01b,Varn13a} The second challenge may seem perhaps more
ascetic, but the objection is firmly grounded in the difficulty that arises
when using the approach to compare experimentally obtained DPs with calculated
ones; as is necessary in many estimation algorithms, such as reverse Monte
Carlo modeling,~\cite{Keen90a} and differential evolution and genetic
algorithms.~\cite{Mich13a}

There are additional advantages to expressing the DP as an \emph{analytical}
function of the HMM. For example, if analytical expressions for the DP of an
arbitrary HMM become available, then a systematic search of \emph{all} HMM
architectures up to a given number of states becomes
feasible.~\cite{John10b,Stre14a} This makes it possible to infer long-range
structures in the so-called \emph{strictly sofic}
processes~\cite{Paz71a,Badi97a} directly from X-ray diffraction studies.
Simulation studies suggest that the sofic stacking processes, which possess a
kind of infinite memory and hence cannot be represented by any finite-order
Markov model, may be important in solid-state transformations.~\cite{Varn04a}
More broadly, we expect that exploring analytical methods will lead to
expressions for material properties, such as electrical conductivities and band
structures, in terms of HMM parameters. This then opens the way to efficiently
and methodically surveying the range of possible (disordered) stacking
processes for those that may result in materials with novel, technologically
useful properties.

In the following, we offer a general, analytical solution to calculating the DP
for stacking processes in CPSs describable as a HMM, thus avoiding the
difficulties inherent in considering sample sequences of a finite length.
Although we specialize to the CPS case, generalizing to other ML geometries and
stacking constraints is straightforward. For our starting point, we assume that
a statistical model of the stacking process is available and given in the form
of a HMM and, from this HMM, we derive analytical expressions for the DPs.
These expressions are valid for any kind and amount of disorder present and,
thus, they encompass virtually all models of disorder in layered CPSs that have
been studied to date. As such, they represent a quite general solution to the
problem. There are, however, processes that are too sophisticated to be
represented by a finite-state HMM, such as the Thue-Morse sequence and the
Fibonacci sequence,~\cite{Baak12a} and they are excluded from this treatment.

Our development is organized as follows: 
\S \ref{Theory} introduces nomenclature and definitions;
\S \ref{DiffPatt} derives a general expression for the DP of layered CPSs in terms of the HMM;
\S \ref{Examples} considers several examples, namely
(i) an independently distributed process that can model 3C or random stacking structures, 
(ii) an infinite-Markov-order stacking process that represents any amount of
random growth and deformation faults in 3C and 2H structures, and 
(iii) a stacking process inspired by recent experiments in 6H-SiC; and \S
\ref{Conclusions} gives our conclusions and outline directions for future work.

\section{Definitions and Notations}
\label{Theory}

Let us make the following assumptions concerning the stacking of MLs in CPSs:
\begin{itemize}
\setlength{\topsep}{0mm}
\setlength{\itemsep}{0mm}
\item The MLs themselves are undefected and free of any distortions;
\item the spacing between MLs does not depend on the local stacking arrangement;
\item each ML has the same scattering power; and
\item the faults extend laterally completely across the crystal.
\end{itemize}
Additionally, we assume that the unconditioned probability of finding a given
stacking sequence remains constant through the crystal. (In statistics
parlance, we assume that the stacking process is \emph{weak-sense stationary}.
Physically, the process is spatial-translation invariant.)

\subsection{Correlation Functions and Stacking Notation}
\label{SubSect:CorrFunc}

In CPSs, each ML may assume one of three possible orientations, usually labeled
$A$, $B$, and $C$.~\cite{Ashc76a} We say that two MLs in a sequence are
\emph{cyclically} related if the ML further along in the sequence can be
obtained from the earlier ML via a cyclic permutation (\ie, $A \to B \to C \to
A$), and \emph{anticyclically} related if it can be obtained via an anticyclic
permutation (\ie, $A \to C \to B \to A$). It is convenient to introduce three
statistical quantities,~\cite{Varn13a} $\Qcn, \Qan$, and $\Qsn$: the pairwise
(auto)correlation functions (CFs) between MLs that are the probability any two
MLs at separation $n$ are related cyclically (c), anticyclically (a), or have
the same orientation (s), respectively. It is also useful to introduce a family
of cyclic-relation functions~\cite{Riec14b} ${\xoperator} (\symbolABC) \in
\{\coperator(\symbolABC), \aoperator(\symbolABC), \soperator(\symbolABC)\}$,
where, for example: 
\begin{equation}
  \coperator(\symbolABC) = \left\{ 
  \begin{array}{l l}
    B & \quad \textrm{if } \symbolABC = A\\
    C & \quad \textrm{if } \symbolABC = B\\
    A & \quad \textrm{if } \symbolABC = C\\
  \end{array} \right..
\end{equation}
The other two operators, $\aoperator(\symbolABC)$ and $\soperator(\symbolABC)$,
are defined in an obviously analogous fashion.

It is sometimes advantageous to exploit the constraint that no two adjacent MLs
may occupy the same orientation in CPSs. Thus, we sometimes use the
\Hagg-notation, where cyclic transitions between adjacent MLs are denoted with
`+', and anticyclic ones with `-'. (Ortiz \etal~\cite{Orti13a} give an
excellent treatment of the various notations used to describe CPSs.) Often it
is more convenient to substitute `1' for `+' and `0' for `-' and make this
substitution throughout. The two notations, the \Hagg-notation and the
$ABC$-notation, carry an equivalent message (up to an overall rotation of the
specimen about the stacking direction), albeit in different tongues. 

\subsection{The Stacking Process as a HMM}
\label{StackingProcessHMM}

Previously, it was shown that the stacking process for many cases of practical
interest can be written as a discrete-step, discrete-state HMM,~\cite{Riec14b}
and we review notations and conventions now.

We assume that the statistics of the stacking process are known and can be
expressed as a HMM in the form of an ordered tuple $\Machine = (\Alphabet,
\States, \mu_0, \TransMatSet)$, where $\Alphabet$ is a set of symbols output by
the process and often called an \emph{alphabet}, $\States$ is a finite set of
$\Nstates$ internal (and possibly hidden) states, $\mu_0$ is an initial state
probability distribution, and $\TransMatSet$ is a set of $|\Alphabet|$
$\Nstates$-by-$\Nstates$ transition matrices (TMs) that give the transition
probabilities between states on emission of one of the symbols in $\Alphabet$.

For the CPSs, the output symbols are just ML orientations and, thus, this
alphabet can either be written in the \Hagg-notation or the $ABC$-notation.
Since the latter is more convenient for our purposes, we take $\Alphabet =
\AlphabetABC \equiv \{A,B,C\}$. $\States$ is the set of $\Nstates$ states that
comprise the process; \ie, $\States = \{\StateABC_1, \StateABC_2, \ldots,
\StateABC_{\Nstates}\}$. Lastly, there is one $\Nstates \times \Nstates$ TM for
each output symbol, so that $\TransMatSet =
\{\TransMatABC^{[A]},\TransMatABC^{[B]},\TransMatABC^{[C]}\}$. These
emission-labeled transition probability matrices are of the form:
\begin{align}
  \TransMat^{[\symbolABC]} =
  \begin{bmatrix}  \nonumber
  \Pr(\symbolABC, \StateABC_1|\StateABC_1 ) \! & \! \Pr(\symbolABC, \StateABC_2|\StateABC_1 ) \! & \! \cdots \! & \! \Pr(\symbolABC, \StateABC_M|\StateABC_1 ) \\
  \Pr(\symbolABC, \StateABC_1|\StateABC_2 ) \! & \! \Pr(\symbolABC,
  \StateABC_2|\StateABC_2 ) \! & \! \cdots \! & \! \Pr(\symbolABC, \StateABC_M|\StateABC_2 ) \\
  \vdots  \! & \! \vdots  \! & \! \ddots \! & \! \vdots  \\
  \Pr(\symbolABC, \StateABC_1|\StateABC_M ) \! & \! \Pr(\symbolABC,
  \StateABC_2|\StateABC_M ) \! & \! \cdots \! & \! \Pr(\symbolABC, \StateABC_M|\StateABC_M ) \\
  \end{bmatrix} ,
\end{align}
where $\symbolABC \in \AlphabetABC$ and $\StateABC_1, \StateABC_2, \dots, \StateABC_M \in \States$.

It is often useful to have the total state-to-state TM, whose components are
the probability of transitions independent of the output symbol, and it is
given by the row-stochastic matrix $\TransMatABC = \TransMatABC^{[A]}
+\TransMatABC^{[B]} + \TransMatABC^{[C]}$. There also exists a stationary
distribution $\mathbf{\Dist} = \big(\Pr(\StateABC_1), \ldots, \Pr(\StateABC_M)
\big)$ over the hidden states, such that $\bra{\Dist} = \bra{\Dist}
\TransMatABC$. We make limited use of a bra-ket notation throughout the
following, where bras $\bra{\cdot}$ represent row vectors and kets
$\ket{\cdot}$ represent column vectors. Bra-ket closures, $\braket{ \cdot}$ or
$\braket{ \cdot | \cdot }$, are scalars and commute as a unit with anything.

In the \Hagg\ representation, the state-to-state transition matrix is $\HaggT =
\HaggTzero + \HaggTone$. In that case the stationary distribution
$\Dist_{\textrm{H}}$ can be obtained from $\bra{\Dist_{\textrm{H}}} =
\bra{\Dist_{\textrm{H}}} \HaggT$.

HMMs are often depicted as labeled directed graphs called probabilistic
finite-state automata (FSA).~\cite{Hopc79a,Paz71a} When written using the
$ABC$-notation, we refer to such an automaton as the \MachineABC\ and,
similarly, when written in terms of the \Hagg-notation, such an automaton is
referred to as the \MachineHagg. It is a straightforward task to translate a
\MachineHagg\ into an \MachineABC.~\cite{Riec14b} For completeness, we
reproduce the minimal algorithm in the appendix.

We note that while \Hagg-notation and \MachineHagg s are useful shorthand, the
primary mathematical object for the development here is the \MachineABC, since
this describes the stacking process in the natural language of the $\{ \Qxin
\}$. It is, however, often easier to give just the \MachineHagg\ since the
expansion procedure is straightforward. Fundamentally, however, it is the $ABC$
sequences that directly relate to structure factors for the specimen. And, this
practical consideration is the principle reason for using the $ABC$-notation
and \MachineABC s.

\subsection{Mixing and Nonmixing Machines}
\label{MixingNonmixingMachines}

When expanding the \MachineHagg\ into an \MachineABC, two important cases
emerge: \emph{mixing} and \emph{nonmixing} \MachineHagg s. Which of these two
cases we are considering has implications for the resultant DP, and so it is
important to distinguish them.~\cite{Riec14b}

In the expansion process, the number of states is tripled to account for the
possible degeneracy of the $ABC$-notation. That is, we require that the
\MachineABC\ keep track of not only the relative orientation between adjacent
MLs (as the \MachineHagg\ does), but also the absolute $A$, $B$, or $C$
orientation. In doing so, we allow a state architecture that can accommodate
this increased representation requirement. For \emph{mixing} machines, the
resultant FSA is \emph{strongly connected}, such that any state is accessible
to any other state in a finite number of transitions. We find that this is by
far the more common case. For \emph{nonmixing} machines, the resultant graph is
not strongly connected, but instead breaks into three unconnected graphs, each
retaining the state structure of the original \MachineHagg. Only one of these
graphs is physically realized in any given specimen, and we may arbitrarily
choose to treat just one of them. The deciding factor on whether a machine is
mixing or nonmixing depends on its architecture: if there exists at least one
closed, nonself-intersecting state path that corresponds to an overall rotation
of the specimen, then the machine is mixing. The closed path is called a
\emph{simple state cycle} (SSC) on a FSA or a \emph{causal state cycle} (CSC)
if the FSA is also an \eM. All of the examples we consider here are mixing
machines over most of their parameter range.

\subsection{Power Spectra}
\label{PowerSpectra}

Since we are considering only finite-state HMMs, $\abcT$ is a
finite-dimensional square matrix, and so its spectrum is just its set of eigenvalues: 
\begin{align}
\Lambda_{\abcT} = \{ \lambda \in \mathbb{C}: \text{det}(\lambda
\IdentMat - \abcT) = 0 \}~,
\end{align}
where $\IdentMat$ is the $\Nstates \times \Nstates$ identity matrix. Since
$\abcT$ is row-stochastic (\ie, all rows sum to one), all of its eigenvalues
live on or within the unit circle in the complex plane. The connection between
the operator's spectrum and the diffraction spectrum will become clear shortly.
In brief, though, eigenvalues along the unit circle lead to Bragg peaks; eigenvalues within the unit circle are responsible for diffuse peaks associated with disorder---the diffuse DP is the shadow that these eigen-contributions cast
along the unit circle.

In the limit of infinite-length sequences,\footnote{For power spectra of
finite-length sequences, there is no clear distinction among these features.}
power spectra generally can be thought of as having three different
contributions; namely, \emph{pure point} ($pp$), \emph{absolutely continuous}
($ac$), and \emph{singular continuous} ($sc$). Thus, a typical power spectrum
${\mathcal {P}}(\omega)$ can be decomposed into:\cite{Badi97a}
\begin{align}
  {\mathcal {P}}(\omega) & = {\mathcal {P}}_{pp}(\omega)
  + {\mathcal {P}}_{ac}(\omega) + {\mathcal {P}}_{sc}(\omega)
  ~.
\end{align}
Pure point spectra are physically realized as Bragg reflections in DPs, and
diffuse or broadband scattering is associated with the absolutely continuous
part. Singular continuous spectra are not often observed in DPs from
quasi-one-dimensional crystals, although specimens can be engineered to have a
singular continuous portion in the DP, as for example layered GaAs-AlAs
heterostructures stacked according to the Thue-Morse process.~\cite{Axel91a}
Since these more exotic processes are not expressible as finite-state HMMs, we
do not consider them further for now.

It might be thought that the more pedestrian forms of disorder in layered
materials---such as growth, deformation, or layer-displacement faults---always
destroy the long-range periodicity along the stacking direction and, thus,
`true' Bragg reflections need not be treated. (This is in contrast to those
cases where there is little disorder and the integrity of the Bragg reflections
is largely preserved.) In fact, there are occasions, such as solid-state
transformations in materials with competing interactions between
MLs~\cite{Kabr88a,Seba94a} or those with disordered and degenerate ground
states~\cite{Yi96a,Varn01a} that do maintain long-range correlations. Hence, it
is not possible to exclude the existence of Bragg reflections {\it {a priori}}.
Thus, we generally consider both Bragg reflections (${\mathrm {B}}$) and
diffuse scattering (${\mathrm {D}}$) here and write the DP $\DP(\ell)$ as
having two contributions:
\begin{align}
      \DP(\ell) & = \DP_{{\mathrm {B}}}(\ell) + \DP_{{\mathrm {D}}}(\ell)~,
\end{align}
where $\ell \in \mathbb{R}$ is a continuous variable that indexes the magnitude
of the perpendicular component of the diffracted wave: $k = \omega / c = 2 \pi
\ell / c$ with $c$ being the distance between adjacent MLs of the crystal.

Fortunately, knowledge of the HMM allows us to select beforehand those values
of $\ell$ potentially contributing Bragg reflections. Let
$\Lambda_{\rho(\abcT)} \equiv \{ \lambda \in \Lambda_{\abcT} : \left| \lambda
\right| = 1 \}$. The values of $\ell$ for which $\e^{i 2 \pi \ell} \in
\Lambda_{\rho(\abcT)}$ are the only ones where there may possibly exist Bragg
reflections. It is immediately apparent, then, that the total number of Bragg
reflections within a unit interval of $\ell$ in the DP cannot be more than the
number $\Nstates$ of HMM states. Conversely, the total number of Bragg
reflections sets a minimum on $\Nstates$.

\section{Diffraction Patterns from HMMS}
\label{DiffPatt}

With definitions and notations in place, we now derive our main results:
analytical expressions for the DP in terms of the parameters that define a
given HMM. We split our treatment into two steps: (i) we first treat the
diffuse part of the spectrum and, then, (ii) we treat those $z$-values ($z
\equiv \e^{i \omega} = \e^{i 2 \pi \ell}$) corresponding to eigenvalues of the
TM along the unit circle.

\subsection{Diffuse Scattering}
\label{DiffuseScattering}

The corrected DP~\footnote{As previously done,~\cite{Varn13a} we
divide out those factors associated with experimental corrections to
the observed DP, as well as the total number $\NModLayers$ of MLs, so that
$\DP(\ell)$ has only those contributions arising from the stacking structure
itself. Here and elsewhere, we refer to $\DP(\ell)$ simply as the DP.} for CPSs
along a row defined by $h_0-k_0 = 1 \pmod{3}$, where $h_0,k_0$ are components
of the reciprocal lattices vectors in the plane of the MLs, can be written
as:~\cite{Yi96a,Este03a,Varn13a}
\begin{widetext}
\begin{align}
\DP^{(\NModLayers)}(\ell) & =
  \frac{\sin^{2}( \NModLayers \pi  \ell )}{\NModLayers\sin^{2}(\pi \ell)}
  -\frac{2\sqrt{3}}{\NModLayers}
  \sum_{n=1}^{\NModLayers} (\NModLayers -n)
  \Bigl[ \Qcn \cos \left( 2\pi n\ell+\tfrac{\pi}{6} \right) 
  + \Qan \cos \left( 2 \pi n\ell - \tfrac{\pi}{6} \right) \Bigr]
\label{eq:diff1} \\ 
& = \frac{\sin^{2}( \NModLayers \pi \ell )}{\NModLayers\sin^{2}(\pi \ell)}
  -\frac{2\sqrt{3}}{\NModLayers} \, \realpart \left\{ \sum_{n=1}^{\NModLayers} (\NModLayers -n) \left[ \Qcn \, \e^{-i 2 \pi n \ell} \e^{-i \pi /6 } 
  + \Qan \, \e^{-i 2 \pi n \ell} \e^{i \pi / 6} \right] \right\} ~.
\end{align}
$\Qcn$ and $\Qan$ are the previously defined CFs and $\NModLayers$ is the total
number of MLs in the specimen.~\footnote{It may seem that specializing to such
a specific expression for the DP at this stage limits the applicability of
the approach. While the development here is restricted to the case of CPS,
under mild conditions, the Wiener-Khinchin theorem~\cite{Badi97a} guarantees
that power spectra can be written in terms of pair autocorrelation functions,
as is done here. This makes the spectral decomposition rather generic.}
The superscript $\NModLayers$ on $\DP^{(\NModLayers)}(\ell)$ reminds us that
this expression for the diffuse DP depends on the number of MLs. The first term
in Eq.~(\ref{eq:diff1}) is the Fej\'er kernel. As the number of MLs becomes
infinite, this term will tend to a $\delta$-function at integer values of $\ell$,
which may be altered or eliminated by $\delta$-function contributions from the
summation: an issue we address shortly. It is only the second term, the
summation, that results in diffuse scattering even as $N \to \infty$. It has
previously been shown~\cite{Riec14b} that the CFs, in turn, can be written in
terms of the labeled and unlabeled TMs of the underlying stacking process as:
\begin{align}
\Qxin &= \sum_{\symbolABC \in \AlphabetABC}
   \bra \Dist \TransMatABC^{[\symbolABC]} \TransMatABC^{n-1} 
             \TransMatABC^{[\xoperator (\symbolABC)]} \ket \One
  ~,
\label{eq:CorrFunc1}
\end{align}
where we denote the asymptotic probability distribution over the HMM states as
the length-$\Nstates$ row vector $\bra \Dist$ and a length-$\Nstates$ column
vector of 1s as $\ket \DistOne$. For mixing processes, Eq.~\eqref{eq:CorrFunc1}
simplifies to the more restricted set of equations:
\begin{align}
       \Qxin = 3 \bra \Dist  \TransMatABC^{[\symbolABC]} \TransMatABC^{n-1} 
       \TransMatABC^{[\xoperator (\symbolABC)]} \ket \One , \text{ where } \symbolABC \in \AlphabetABC~.
\label{eq:CorrFunc2}       
\end{align}
Thus, we can rewrite the DP directly in terms of the TMs of the underlying
stacking process as:
\begin{align}
\DP^{(\NModLayers)}(\ell) = &  
\frac{\sin^{2}( \NModLayers \pi \ell )}{\NModLayers\sin^{2}(\pi \ell)}
   -\frac{2\sqrt{3}}{\NModLayers} \, \realpart 
   \left\{ \sum_{\symbolABC \in \AlphabetABC}
   \bra \Dist \TransMatABC^{[\symbolABC]} 
   \left( \sum_{n = 1}^\NModLayers
   \left( \NModLayers - n \right)  z^{-n}  \TransMatABC^{n-1} \right)
   \left( \e^{-i\pi/6 } \TransMatABC^{[\coperator(\symbolABC)]}
   + \e^{i \pi / 6 } \TransMatABC^{[\aoperator(\symbolABC)]}\right)
   \ket \One \right\}
   ~,
\label{eq:DiffuseDiffPatt1}
\end{align}
where we have introduced the $\ell$-dependent variable $z \equiv \e^{i 2 \pi
\ell }$. Furthermore, we can evaluate the summation over $n$ in
Eq.~(\ref{eq:DiffuseDiffPatt1}) analytically. First, we note that the summation
can be re-indexed and split up as:
\begin{align}
\sum_{n = 1}^\NModLayers \left( \NModLayers - n \right) z^{-n} \abcT^{n-1}
  & = z^{-1} \sum_{\eta = 0}^{\NModLayers-1}
  \left( \NModLayers - 1 - \eta \right)  \left( \abcT / z \right)^{\eta} \\
  & = z^{-1} \left\{ (\NModLayers -1)
  \left[ \sum_{\eta = 0}^{\NModLayers - 1} \left( \abcT / z \right)^{\eta} \right]
  - \left[ \sum_{\eta = 0}^{\NModLayers - 1} \eta \left( \abcT / z \right)^{\eta} \right]
\right\}
  ~.
\end{align}
For finite positive integer $N$, it is always true that:
\begin{align}
(z \IdentMat - \abcT)
  \sum_{\eta = 0}^{\NModLayers - 1} \left( \abcT / z \right)^{\eta}
  = z \left[ \IdentMat - ( \abcT / z )^{\NModLayers} \right]
\end{align}
and
\begin{align}
(z \IdentMat - \abcT)
  \sum_{\eta = 0}^{\NModLayers - 1} \eta \left( \abcT / z \right)^{\eta}
  = z \left\{ \left[ \sum_{\eta = 0}^{\NModLayers - 1}
  \left( \abcT / z \right)^{\eta} \right] 
  - N ( \abcT / z )^{\NModLayers} 
  - \left[ \IdentMat - ( \abcT / z )^{\NModLayers} \right] \right\}
  ~.
\end{align}
Hence, for $z \notin \Lambda_{\abcT}$, $z \IdentMat - \abcT$ is invertible and we have:
\begin{align}
\sum_{n = 1}^{\NModLayers} \left({\NModLayers} - n \right) z^{-n} \abcT^{n-1}
  & = ( z \IdentMat - \abcT )^{-1}
  \left\{ \NModLayers \IdentMat - z (z \IdentMat - \abcT)^{-1} 
  \left[ \IdentMat - ( \abcT / z )^{\NModLayers} \right]  \right\}
  `.
\end{align}
Putting this all together, we find the expected value of the finite-$N$ DP for
all $z = \e^{i 2 \pi \ell } \notin \Lambda_{\abcT}$:
\begin{align}
\DP^{(\NModLayers)} & (\ell) =
\frac{\sin^{2}( \NModLayers \pi \ell )}{\NModLayers\sin^{2}(\pi \ell)}
  \nonumber \\
  & -2\sqrt{3} \, \realpart \left\{ \sum_{\symbolABC \in \AlphabetABC}
  \bra \Dist \TransMatABC^{[\symbolABC]} 
  ( z \IdentMat - \abcT )^{-1}
  \left\{ \IdentMat - \frac{z}{\NModLayers} (z \IdentMat - \abcT)^{-1} 
  \left[ \IdentMat - ( \abcT / z )^{\NModLayers} \right]  \right\}
  \left( \e^{-i\pi/6 } \TransMatABC^{[\coperator(\symbolABC)]}
  + \e^{i \pi / 6 } \TransMatABC^{[\aoperator(\symbolABC)]}\right)
  \ket \One \right\}
  ,
\label{eq:DiffuseDiffPatt2}
\end{align}
with $z \equiv e^{i 2 \pi \ell }$. This gives the most general relationship
between the DP and the TMs of the underlying stacking process. We see that the
effects of finite crystal size come into the diffuse DP via a
$1/\NModLayers$-decaying term containing the $\NModLayers^{th}$ power of both
$z^{-1}$ and the unlabeled TM. This powerful result directly links the stacking
process rules to the observed DP and, additionally, already includes the
effects of finite specimen size.

For many cases of practical interest, the specimen can be treated as
effectively infinite along the stacking direction. (In follow-on work, we
explore the effects of finite specimen size.) In this limiting case, the
relationship between the diffuse DP and the TMs becomes especially simple.  In
particular, as $\NModLayers \to \infty$ the DP's diffuse part becomes:
\begin{align}
\DP_{{\mathrm {D}}}(\ell) 
  & = \lim_{\NModLayers \to \infty} \DP^{(\NModLayers)}(\ell) 
  = - 2 \sqrt{3} \realpart \left\{ \sum_{\symbolABC \in \AlphabetABC}
  \bra \Dist  \TransMatABC^{[\symbolABC]} 
  \left( z \IdentMat - \TransMatABC \right)^{-1}
  \left( \e^{- i \pi / 6 } \TransMatABC^{[\coperator(\symbolABC)]}
  + \e^{ i \pi / 6 } \TransMatABC^{[\aoperator(\symbolABC)]}  \right) 
  \ket \One \right\} 
  ~,
\label{eq:DiffuseDiffPatt4}
\end{align}
for all $z = \e^{i 2 \pi \ell } \notin \Lambda_{\abcT}$. For mixing processes,
this reduces to: 
\begin{align}
\DP_{{\mathrm {D}}}(\ell) 
  & = - 6 \sqrt{3} \realpart \left\{
  \bra \Dist \TransMatABC^{[\symbolABC]} 
  \left( z \IdentMat - \TransMatABC \right)^{-1}
  \left( \e^{- i \pi / 6 } \TransMatABC^{[\coperator(\symbolABC)]}
  + \e^{ i \pi / 6 } \TransMatABC^{[\aoperator(\symbolABC)]}  \right) 
  \ket \One \right\} 
  ~,
\label{eq:MixingDiffuseDP}
\end{align}
\end{widetext}
for any $\symbolABC \in \AlphabetABC$. Note that there are no powers of the TM
that need to be calculated in either of these cases. Rather, \emph{the DP is a
direct fingerprint of the noniterated TMs}. The simple elegance of
Eq.~(\ref{eq:DiffuseDiffPatt4}) relating the DP and TMs suggests that there is
a link of fundamental conceptual importance between them. The examples to
follow draw out this connection.

The important role that $\abcT$'s eigenvalues $\Lambda_{\abcT}$ play in the DP
should now be clear: they are the poles of the resolvent matrix $(\zeta
\IdentMat - \abcT)^{-1}$ with $\zeta \in \mathbb{C}$. Since the DP is a simple
function of the resolvent evaluated along the unit circle, $\Lambda_{\abcT}$
plays a critical organizational role in the DP's structure. Any peaks in the DP
are shadows of the poles of the resolvent filtered through the appropriate row
and column vectors and cast out radially onto the unit circle. Peaks in the DP
become more diffuse as the corresponding eigenvalues withdraw towards the
origin of the complex plane. They approach $\delta$-functions as the corresponding
eigenvalues approach the unit circle. \S \ref{Examples}'s examples demonstrate
this graphically.

\subsection{Bragg Reflections}
\label{BraggReflections}

The eigenvalues $\Lambda_{\rho(\abcT)} \subset \Lambda_{\abcT}$ along the
unit circle are responsible for Bragg peaks, and we treat this case now.  For finite-$N$, the eigenvalues along the unit circle give rise to Dirichlet kernels. As $N \to \infty$, the analysis becomes somewhat simpler since the Dirichlet kernel and Fej\'er kernel both tend to $\delta$-functions.

In the limit of $\NModLayers \to \infty$, the summation over $n$ in Eq.~(\ref{eq:DiffuseDiffPatt1})
divided by the total number of modular layers becomes:
\begin{align}
\lim_{\NModLayers \to \infty} \sum_{n = 1}^\NModLayers \frac{\NModLayers - n}{\NModLayers} z^{-n} \abcT^{n-1}
&=
z^{-1} \sum_{\eta = 0}^{\infty} \left( \abcT / z \right)^{\eta}~.
\label{eq:InftSumKernel}
\end{align}
At this point, it is pertinent to use the spectral decomposition of $\abcT^L$,
developed by us.~\cite{Riec14a} With the allowance that $0^{L-m} =
\delta_{L-m,0}$ for the case that $0 \in \Lambda_{\abcT}$, this is:
\begin{align}
\abcT^L 
& = \sum_{\lambda \in \Lambda_{\abcT} } 
		\sum_{m=0}^{\nu_\lambda - 1}  
		\lambda^{L-m}  \binom{L}{m} 
		\abcT_\lambda  
		\left( \abcT - \lambda \IdentMat \right)^m
  ~,
\label{eq: T^L spectral decomp for positive integer L}		
\end{align}
where (i) $\abcT_\lambda$ is the projection operator associated with the
eigenvalue $\lambda$ given by the elementwise residue of the resolvent $\left(
z \IdentMat - \abcT \right)^{-1}$ at $z \to \lambda$, (ii) the index
$\nu_\lambda$ of the eigenvalue $\lambda$ is the size of the largest Jordan
block associated with $\lambda$, and (iii) $\binom{L}{m} = \frac{1}{m!}
\prod_{n=1}^m (L-n+1)$ is the generalized binomial coefficient. In terms of
elementwise counter-clockwise contour integration, we have:
\begin{align}
\abcT_\lambda = \frac{1}{2 \pi i} \oint_{C_\lambda}
  \left( z \IdentMat - \abcT \right)^{-1} \, dz
  ~, 
\label{eq:GeneralProjOpEqn}
\end{align}
where $C_\lambda$ is any contour in the complex plane enclosing the point $z_0
= \lambda$---which may or may not be a singularity depending on the particular
element of the resolvent matrix---but encloses no other singularities.
Usefully, the projection operators are a mutually orthogonal set such that for
$\zeta, \lambda \in \Lambda_{\abcT}$, we have:
\begin{align*}
\abcT_\zeta \abcT_\lambda = \delta_{\zeta, \lambda} \abcT_\lambda
  ~.
\end{align*}

The Perron--Frobenius theorem guarantees that all eigenvalues of the stochastic
TM $\abcT$ lie on or within the unit circle. Moreover---and very important to
our discussion on Bragg reflections---the eigenvalues on the unit circle are
guaranteed to have an index of one. The indices of all other eigenvalues must
be less than or equal to one more than the difference between their algebraic
$a_\lambda$ and geometric $g_\lambda$ multiplicities. Specifically:
\begin{align*}
\nu_\lambda - 1 \leq a_\lambda - g_\lambda \leq a_\lambda - 1 
\end{align*}
and
\begin{align*}
\nu_\lambda = 1 \, \text{ if } |\lambda| = 1
  ~.
\end{align*}
Taking advantage of the index-one nature of the eigenvalues on the unit circle, 
we can define:
\begin{align*}
\Xi \equiv \sum_{\zeta \in \Lambda_{\rho(\abcT)}} \zeta \abcT_{\zeta}
\end{align*}
and 
\begin{align*}
F \equiv \abcT - \Xi
  ~. 
\end{align*}
Then, the summation on the right-hand side of Eq.~\eqref{eq:InftSumKernel}
becomes:
\begin{align}
\sum_{\eta = 0}^{\infty} \left( \abcT / z \right)^{\eta} & = 
\left[ \sum_{\eta = 0}^{\infty} \left( F / z \right)^{\eta} \right] + 
\left[ \sum_{\eta = 0}^{\infty} \left( \Xi / z \right)^{\eta} \right]~.
\end{align}
In the above, only the summation involving $\Xi$ is capable of contributing
$\delta$-functions. And so, expanding this summation, yields:
\begin{align}
\sum_{\eta = 0}^{\infty} \left( \Xi / z \right)^{\eta} & = 
\sum_{\lambda \in \Lambda_{\rho(\abcT)}} \abcT_{\lambda} 
\sum_{\eta = 0}^{\infty} \left( \lambda / z \right)^{\eta} \\
& = 
\sum_{\lambda \in \Lambda_{\rho(\abcT)}} \abcT_{\lambda} 
\sum_{\eta = 0}^{\infty} \e^{ i 2 \pi \left( \ell_\lambda - \ell \right) \eta}~, 
\label{eq:SummationWExponential}
\end{align}
where $\ell_\lambda$ is related to $\lambda$ by $\lambda = \e^{i 2 \pi \ell_\lambda}$
over some appropriate length-one $\ell$-interval. 

Using properties of the discrete-time Fourier transform (DTFT),~\cite{Oppe75a} we can finally pull the $\delta$-functions out of Eq.~\eqref{eq:SummationWExponential}. In particular: 
\begin{align}
\sum_{\eta = 0}^{\infty} & \e^{ i 2 \pi \left( \ell_\lambda - \ell \right) \eta}
  \nonumber \\
  & = \frac{1}{1 - \e^{i 2 \pi (\ell_\lambda - \ell)}}
  + \sum_{k = -\infty}^{\infty} \tfrac{1}{2} \delta(\ell - \ell_\lambda + k)
  ~.
\label{eq:DiracComb}
\end{align}

Identifying the context of Eq.~\eqref{eq:DiracComb} within
Eq.~\eqref{eq:DiffuseDiffPatt1} shows that the potential $\delta$-function at
$\ell_\lambda$ (and at its integer-offset values) has magnitude:\footnote{By
magnitude, we mean the $\ell$-integral over the $\delta$-function. If integrating with respect to a related variable, then the magnitude of the $\delta$-function changes accordingly.  As a simple example, integrating over $\omega = 2\pi \ell$ changes the magnitude of the $\delta$-function by a factor of $2 \pi$.}
\begin{align}
\Delta_{\lambda} & \equiv 
  \lim_{\epsilon \to 0 } \int_{\ell_{\lambda} - \epsilon}^{\ell_{\lambda}
  + \epsilon} \DP(\ell) \, d\ell \nonumber \\ 
  & = - \sqrt{3} \, \realpart \left\{ \lambda^{-1} \left[ 
  \Braket{\abcT_\lambda ^{\coperator(\Alphabet)}} \e^{- i \pi / 6 } + 
  \Braket{\abcT_\lambda ^{\aoperator(\Alphabet)}} \e^{ i \pi / 6 }
  \right]
  \right\}
\label{eq:DeltaFnMagnitudes}
\end{align}
contributed via the summation of Eq.~\eqref{eq:DiffuseDiffPatt1}, where:
\begin{align}
\CFBraKet \equiv 
\sum_{\symbolABC_{0} \in \AlphabetABC} \bra \Dist 
        \TransMatABC^{[\symbolABC_{0}]} \abcT_\lambda  
        \TransMatABC^{[\xoperator (\symbolABC_{0})]} \ket \One
  ~.
\label{eq:DiagonalizableCorrelationAmplitudes}
\end{align}

Finally, considering Eq.~\eqref{eq:DeltaFnMagnitudes} together with the contribution of the 
persistent Fej\'er kernel, the discrete part of the DP is given by:
\begin{align}
\DP_{{\mathrm {B}}}(\ell) & =
\sum_{k = -\infty}^{\infty}  
\sum_{\lambda \in \Lambda_{\rho(\abcT)} } 
\left( \delta_{\lambda, 1} +  \Delta_\lambda \right) \delta(\ell - \ell_\lambda + k)~,
\label{eq:DiscreteDP}
\end{align}
where $\delta_{\lambda, 1}$ is a Kronecker delta and $\delta(\ell - \ell_\lambda + k)$ is a 
Dirac $\delta$-function. 

In particular, the presence of the Bragg reflection at integer $\ell$ (zero
frequency) depends strongly on whether the stacking process is mixing. In any
case, the magnitude of these $\delta$-functions at integer $\ell$ is $1 +
\Delta_1$. For an ergodic process $\abcT_1 = \ket \One \bra \Dist$, so we have:
\begin{align}
\Braket{\abcT_1 ^{\xoperator (\Alphabet)}} &=
\sum_{\symbolABC_{0} \in \AlphabetABC} \bra \Dist 
        \TransMatABC^{[\symbolABC_{0}]} \ket \One \bra \Dist
        \TransMatABC^{[\xoperator (\symbolABC_{0})]} \ket \One
  ~.
\label{eq:UnityCorrelationAmplitude}
\end{align}

For \emph{mixing} $ABC$-machines,
$\bra \Dist \TransMatABC^{[\symbolABC]} \ket \One = \Pr(\symbolABC) = 1 / 3$
for all $\symbolABC \in \AlphabetABC$, giving $\Braket{\abcT_1 ^{\xi (\Alphabet)}} = 1/3$. Hence: 
\begin{align}
\Delta_1 & = 
- \frac{\sqrt{3}}{3} \, \realpart \left\{  
 \e^{- i \pi / 6 } +  \e^{ i \pi / 6 }
 \right\} \nonumber \\
& = 
- \frac{2 \sqrt{3}}{3} \cos (\pi / 6) \nonumber \\
& = -1~,
\end{align}
and the integer-$\ell$ $\delta$-functions are extinguished for all mixing processes.

For \emph{nonmixing} processes, the probability of each ML is \emph{not}
necessarily the same, and the magnitude of the $\delta$-function at
integer-$\ell$ will reflect the heterogeneity of the single-symbol statistics.

\subsection{Full Spectral Treatment of the Diffuse Spectrum}

From Eq.~\eqref{eq:DiffuseDiffPatt4}, it is clear that the diffuse part of the
DP is directly related to the resolvent $\left( z \IdentMat - \TransMatABC
\right)^{-1}$ of the state-to-state TM evaluated along the unit circle.
According to Riechers and Crutchfield~\cite{Riec14a} the resolvent can be
expressed in terms of the projection operators:
\begin{align}
\left( z \IdentMat - \TransMatABC \right)^{-1} & = 
\sum_{\lambda \in \Lambda_{\abcT}} \sum_{m=0}^{\nu_\lambda-1} 
    \tfrac{1}{(z-\lambda)^{m+1}} 
    \abcT_\lambda (\abcT - \lambda \IdentMat)^m~.
\end{align}
Hence, Eq.~\eqref{eq:DiffuseDiffPatt4} can be expressed as:
\begin{widetext}
\begin{align}
\DP_{{\mathrm {D}}}(\ell) 
& = 
- 2 \sqrt{3} \, \realpart \left\{ \sum_{\lambda \in \Lambda_{\abcT}} \sum_{m=0}^{\nu_\lambda-1} 
    \frac{1}{(z-\lambda)^{m+1}}  \left[ 
\Braket{\abcT_{\lambda, m} ^{\coperator(\Alphabet)}} \e^{- i \pi / 6 } + 
\Braket{\abcT_{\lambda, m} ^{\aoperator(\Alphabet)}} \e^{ i \pi / 6 }
\right] \right\}~,
\label{eq:SpectralDiffuseDP}
\end{align}
\end{widetext}
where $\CFBraKetm$ is a complex-valued scalar:\footnote{$\CFBraKetm$ is
          constant with respect to the relative layer displacement $n$. However, $\Big\{
          \CFBraKetm \Big\}$ can be a function of a process's parameters.}
\begin{align} 
\CFBraKetm \equiv
\sum_{\symbolABC_{0} \in \AlphabetABC} \bra \Dist 
          \TransMatABC^{[\symbolABC_{0}]} 
          \abcT_\lambda  \left( \abcT - \lambda \IdentMat \right)^m
          \TransMatABC^{[\xoperator (\symbolABC_{0})]} \ket \DistOne
		  ~.
\end{align}
Moreover, if $\Braket{\abcT_{\lambda, m} ^{\coperator(\Alphabet)}} = 
\Braket{\abcT_{\lambda, m} ^{\aoperator(\Alphabet)}}$ for all $\lambda$ and all $m$,
then Eq.~\eqref{eq:SpectralDiffuseDP} simplifies to:
\begin{align}
\DP_{{\mathrm {D}}}(\ell) 
& = 
- 6 \realpart \left\{ \sum_{\lambda \in \Lambda_{\abcT}} \sum_{m=0}^{\nu_\lambda-1} 
    \frac{\Braket{\abcT_{\lambda, m} ^{\coperator(\Alphabet)}}}{(z-\lambda)^{m+1}}  \right\}~.
\label{eq:SymmetricSpectralDiffuseDP}
\end{align}  

\section{Examples}
\label{Examples}

To illustrate the theory, we treat in some detail three examples for which we
previously~\cite{Riec14b} estimated the CFs directly from the HMM. Throughout
the examples, we find it particularly revealing to plot the DP and TM
eigenvalues via, what we call, the \emph{\csg}. This takes advantage of the
fact that the DP is periodic in $\ell$ with period one and that the TM's
eigenvalues lie on or within the unit circle in the complex plane. Thus, a
\csg\ is any frequency-dependent graph emanating radially from the unit circle,
while the unit circle and its interior are concurrently used for its portion of
the complex plane to plot the poles of the resolvent of the underlying
process's transition dynamic.  (Here, the poles of the resolvent are simply the
eigenvalues $\Lambda_\abcT$ of $\abcT$, since $\abcT$ is finite dimensional.) 

\Csg s plot the DP as a function of the polar angle $\omega = 2 \pi \ell$.  The
radial extent of the corona is normalized to have the same maximal value for
each figure here. With our particular interest in the DP of CPSs, we plot all
eigenvalues in $\Lambda_\abcT$ as (red, online) dots and also plot all
eigenvalues in $\Lambda_\HaggT$ as (black) $\times$s. Note that $\Lambda_\HaggT
\subset \Lambda_\abcT$. In all of our examples, it appears that only the
eigenvalues introduced in generating the \MachineABC\ from the \MachineHagg\
(dots without $\times$s through them) are capable of producing DP peaks. For
\emph{nonmixing} processes this is not true, since the \MachineHagg\ and
\MachineABC\ share the same topology and the same set of eigenvalues. 

\subsection{3C Polytypes and Random ML Stacking: IID Process}
\label{RandomStacking}

The independent and identically distributed \Hagg\ process is the simplest ML
stacking process in a CPS that one can consider. Although we work out this
example largely as a pedagogical exercise, in limiting cases it can be thought
of as random deformation faulting in face-center cubic (FCC) (aka 3C) crystals.

\begin{figure}
\begin{center}
\resizebox{!}{3.0in}{\includegraphics{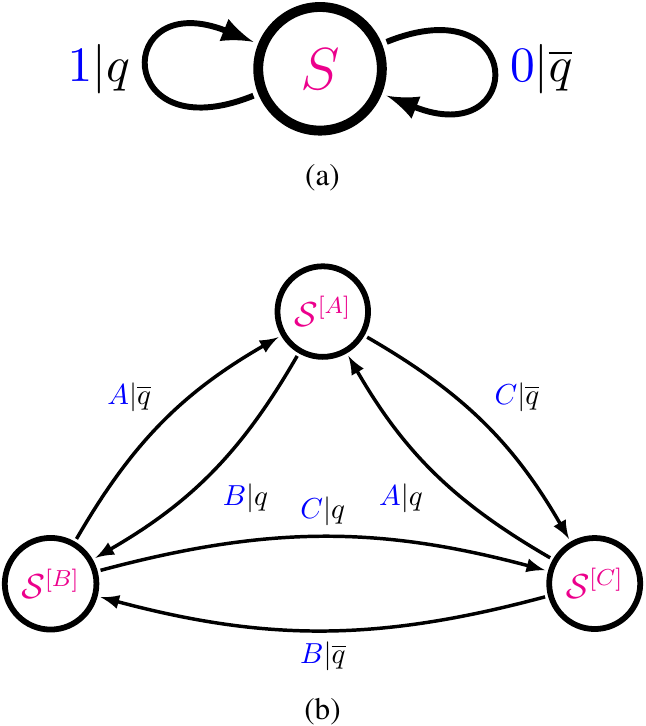}}
\end{center}     
\caption{The (a) \MachineHagg\ and the (b) \MachineABC\ for the IID Process,
  $q \in [0,1]$. When $q=1$, the IID process generates a string of $1$s, which
  is physically the 3C$^+$ stacking structure. Conversely, for $q=0$, the
  structure corresponds to the 3C$^-$ structure. For $q = 0.5$, the MLs are
  stacked as randomly as possible. Notice that the single state of the
  \MachineHagg\ has split into a three-state \MachineABC. This trebling of
  states is a generic feature of expanding mixing \MachineHagg s into
  \MachineABC s.  (From Riechers~\etal, (\citeyear{Riec14b}), used with
  permission.)
  } 
\label{Fig:RandomProcess} 
\end{figure}

We define the \emph{independent and identically distributed} (IID) stacking
process as such: when transitioning between adjacent MLs, a ML will be
cyclically related to the previous ML with probability $q \in [0,1]$. Due to
stacking constraints, the ML will otherwise be anticyclically related to its
predecessor with probability $\bar{q} \equiv 1-q$.\footnote{Here and in the
following examples, we define a bar over a variable to mean one minus that
variable: $\bar{x} \equiv 1-x$.} The \MachineHagg\ and \MachineABC\ for
the IID Process are given in Fig.~\ref{Fig:RandomProcess}.

It is useful to consider limiting cases for $q$. When $q=0.5$, the stacking is completely random, subject only to
the stacking constraints preventing two adjacent MLs from having the same orientation. As $q \to 1$, adjacent
MLs are almost always cyclically related, and the specimen can be thought of as a 3C$^{+}$ crystal with randomly
distributed deformation faults~\cite{Warr69a} with probability $\bar{q}$. As $q \to 0$, it is also a 3C crystal 
with randomly distributed deformation faults, except that the MLs are anticyclically related, which we denote as 3C$^{-}$. 
This is summarized in Table~\ref{Table:RandomFault}.

\begin{table}
\caption{\label{Table:RandomFault} The limiting material structures for the
  IID Process. Key: DF - deformation fault; Ran - completely random stacking.
  }
\begin{ruledtabular}
\begin{tabular}{ccccc}
   $q = 0$  & $q \approx 0$  &  $q = \bar{q} = \frac{1}{2}$  &  $\bar{q} \approx 0$ & $\bar{q} = 0$ \\  \hline
    3C$^{-}$   &   3C$^{-}$/DF    & Ran   &  3C$^{+}$/DF    & 3C$^{+}$  \\
\end{tabular}
\end{ruledtabular}
\end{table}

The TMs in $ABC$-notation are:


\begin{align*}
  \begin{array}{r@{\mskip\thickmuskip}l}
\abcTA
=
\begin{bmatrix}
0 & 0 & 0 \\
\overline{q} & 0 & 0 \\
q & 0 & 0
\end{bmatrix} ,
  \end{array} 
  \begin{array}{r@{\mskip\thickmuskip}l}
\abcTB
=
\begin{bmatrix}
0 & q & 0 \\
0 & 0 & 0 \\
0 & \overline{q} & 0
\end{bmatrix}
          \end{array}
\end{align*}
and
\begin{align}
\abcTC
=
\begin{bmatrix} \nonumber
0 & 0 & \overline{q} \\
0 & 0 & q \\
0 & 0 & 0
\end{bmatrix} .
\end{align}
The internal state TM then is their sum:
\begin{align}
\abcT
=
\begin{bmatrix}  \nonumber
0 & q & \overline{q} \\
\overline{q} & 0 & q \\
q & \overline{q} & 0
\end{bmatrix}
  ~.
\end{align}
The eigenvalues of the $ABC$ TM are 
\begin{align*}
\Lambda_{\abcT} = \{ 1, \, \Omega, \, \Omega^* \}~,
\end{align*}
where:
\begin{align*}
\Omega \equiv - \frac{1}{2} + i \frac{\sqrt{3}}{2} (4q^2 - 4q + 1 ) ^{1/2}
\end{align*}
and $\Omega^*$ is its complex conjugate. 

Furthermore, the stationary distribution over states of the $ABC$-machine can
be found from $\bra \Dist = \bra \Dist \TransMatABC$ to be:
\begin{align*}
    \bra \Dist = \begin{bmatrix} \tfrac{1}{3} & \tfrac{1}{3} & \frac{1}{3} \end{bmatrix}
~.
\end{align*}

\begin{figure*}
\begin{center}
\resizebox{!}{6.0in}{\includegraphics{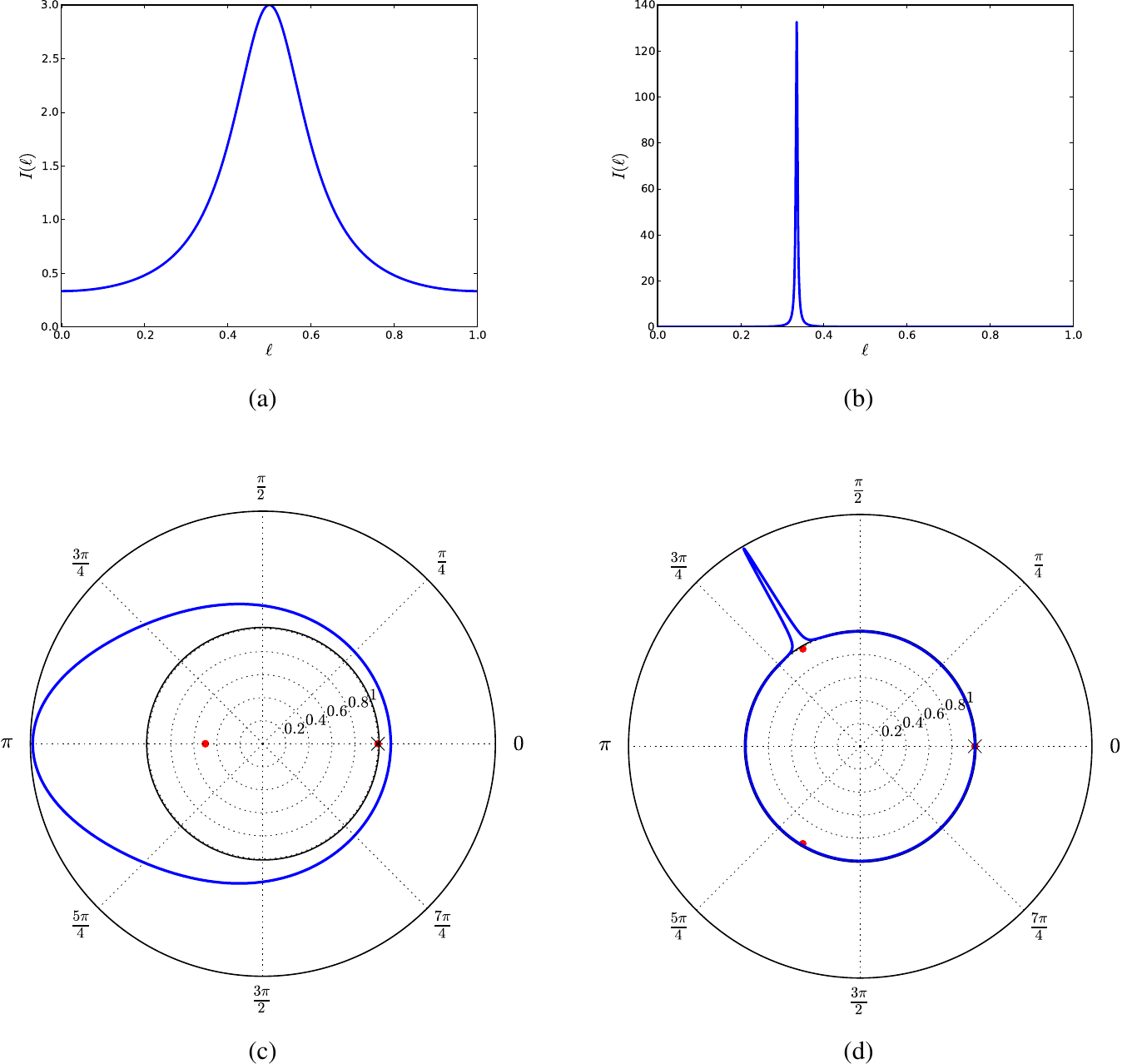}}
\end{center}     
\caption{IID Process diffraction patterns for (a) $q=0.5$ and (b) $q=0.99$, as
  calculated from Eq. (\ref{Eq:IIDProcess}). Notice that as $q \to 1$, the DP
  approaches that of a 3C$^+$ crystal. For values of $q$ close to but less than
  1, the specimen is 3C$^+$ with randomly distributed deformation faults. (c)
  The \csg\ corresponding to $q=0.5$. The enhanced scattering at $\ell = 0.5$
  in (a) is replaced with the bulge at $\omega = \pi$. There are three
  eigenvalues for the IID Process, one at $z=1$ and a degenerate pair at
  $z=-0.5$. (d) The \csg\ corresponding to $q=0.99$. The Bragg-like peak at
  $\ell = 0.33$ in (b) is now represented as a Bragg-like peak at $\omega =
  2\pi/3.$ Notice how the degenerate eigenvalues in (c) have split and migrated
  away from the real axis. As they approach the boundary of the unit circle,
  their presence makes possible Bragg-like reflections in the DP. However,
  eigenvalues near the unit circle are a \emph{necessary}, but \emph{not
  sufficient} condition for Bragg-like reflections. This is seen in the
  eigenvalue in the third quadrant that is \emph{not} accompanied by a
  Bragg-like reflection.
  } 
\label{Fig:IIDCoronalSpectroGrams} 
\end{figure*}

For $q \in (0,1)$, none of the eigenvalues in $\Lambda_{\abcT}$ besides unity
lie on the unit circle of the complex plane, and so there is no possibility of
Bragg reflections at non-integer $\ell$. Moreover, since the process is mixing,
the Bragg peak at integer $\ell$ is also absent. Thus we need only find the
diffuse DP. To calculate the $\DP_{{\mathrm {D}}}(\ell)$ as given in
Eq.~(\ref{eq:DiffuseDiffPatt4}), we are only missing $\left(z \IdentMat -
\TransMatABC \right)^{-1}$, which is given by:
\begin{widetext}
\begin{align*}
\left(z \IdentMat - \TransMatABC \right)^{-1} 
= 
\frac{1}{(z-1)(z - \Omega )(z - \Omega^{*})} 
\times
\begin{bmatrix}
z^2- q\bar{q} 		& qz+ \bar{q}^2  		& \bar{q}z+ q^2  \\
\bar{q}z+ q^2 		& z^2- q\bar{q}			& qz+ \bar{q}^2  \\
qz+ \bar{q}^2  		& \bar{q}z+ q^2 		& z^2 - q\bar{q} 
\end{bmatrix}~.
\end{align*}
Then, with:
\begin{align*}
\bra \Dist \TransMatABC^{[A]} =
\begin{bmatrix}
\tfrac{1}{3} & 0 & 0
\end{bmatrix} ,
\end{align*}
we can write
\begin{align*}
\bra \Dist \TransMatABC^{[A]} \left(z \IdentMat - \TransMatABC \right)^{-1}  =
\frac{1}{3}
\frac{1}{(z-1)(z - \Omega )(z - \Omega^{*})}
& \times
\begin{bmatrix}
z^{2} - q\bar{q}		& qz+ \bar{q}^2 	& \bar{q}z+ q^2 
\end{bmatrix} ,
\end{align*}
\end{widetext}
where:
\begin{align*}
\TransMatABC^{[\coperator(A)]} \ket \One = \TransMatABC^{[B]} \ket \One = 
\begin{bmatrix}
q \\
0 \\
\bar{q} \\
\end{bmatrix} 
\end{align*}
and
\begin{align*}
\TransMatABC^{[\aoperator(A)]} \ket \One = \TransMatABC^{[C]} \ket \One = 
\begin{bmatrix}
\bar{q} \\
q \\
0
\end{bmatrix}
  ~.
\end{align*}
From Eq.~\eqref{eq:MixingDiffuseDP}, the DP becomes:
\begin{align}
\DP_{{\mathrm {D}}}(\ell)
&= 
- \frac{6 \sqrt{3}}{2} \realpart \left\{ 
\frac{e^{i \pi / 6} z \left( q^2 + \bar{q} z \right) + e^{- i \pi / 6} z \left( \bar{q}^2 + q z \right)}{(z-1)(z - \Omega )(z - \Omega^{*})}
\right\} \\
&= 
- 2 \sqrt{3} \realpart \left\{ z \, 
\frac{e^{i \pi / 6}  \left( q^2 + \bar{q} z \right) + e^{- i \pi / 6}  \left( \bar{q}^2 + q z \right)}{(z-1)(z^2 + z + 1-3q \bar{q})}
\right\} .
\label{Eq:IIDProcess}
\end{align}
For the case of the most random possible stacking in CPSs, where $q = \bar{q} = \frac{1}{2}$, this simplifies to: 
\begin{align}
\DP_{{\mathrm {D}}}(\ell)
&=
\frac{3/4}{5/4 + \cos(2 \pi \ell)} .
\end{align}
This result was obtained previously by more elementary means. The results are
in agreement.~\cite{Guin63a}

Figure~\ref{Fig:IIDCoronalSpectroGrams} shows DPs and \csg s for $q=0.5$ and
$q=0.99$. Figure~\ref{Fig:IIDCoronalSpectroGrams}(a) gives the DP for a
maximally disordered stacking process. The spectrum is entirely diffuse with
broadband enhancement near $\ell = 0.5$. In contrast, the DP for $q=0.99$ in
Fig.~\ref{Fig:IIDCoronalSpectroGrams}(b) shows a strong Bragg-like reflection
at $\ell = 0.33$, which we recognize as just the 3C$^+$ stacking structure,
with a small amount of (as it turns out in this case) deformation faulting. The
other two panels in Fig.~\ref{Fig:IIDCoronalSpectroGrams}, (c) and (d), are
\csg s giving DPs for these two cases as the radially emanating curve outside
the unit circle, but now the three eigenvalues of the total TM are plotted
interior to the unit circle. As always, there is a single eigenvalue at $z=1$.
In panel (c), the other two degenerate eigenvalues occur at $z=-0.5$, `casting
a shadow' on the unit circle in the form of enhanced power at $\omega = \pi$.
In panel (d), these eigenvalues split and move away from the real axis closer
to the unit circle. In doing so, one casts a more focused shadow in the form of
a Bragg-like reflection at $\omega = 2\pi/3$. For $q=1$, this eigenvalue
finally comes to rest on the unit circle, and the Bragg-like reflection becomes
a true Bragg peak, as explored shortly. Note that the other eigenvalue does not
give rise to enhanced scattering. We find that having an eigenvalue near the
unit circle is necessary to produce enhanced scattering, but the presence of
such an eigenvalue does not necessarily guarantee Bragg-like reflections.

\subsubsection{Bragg Peaks from 3C}

For the case of $q \in \{ 0, 1 \}$, we recover perfect crystalline structure. 
Although the presence, placement, and magnitude of Bragg peaks are well known
from other methods, we show the comprehensive
consistency of our method via the example of $q=1$ ($\bar{q} = 0$).
In this case:
$\Lambda_{\abcT} = \{ 1, \, \Omega, \, \Omega^* \}$ with 
$\Omega = - \frac{1}{2} + i \frac{\sqrt{3}}{2} = \e^{i 2 \pi /3}$, 
so that $\ell_{\Omega} = 1/3$ and $\ell_{\Omega^*} = 2/3$, 
and the two relevant projection operators reduce to:
\begin{align*}
\abcT_\Omega = \frac{1}{(\Omega-1)(\Omega - \Omega^*)}
\begin{bmatrix}
\Omega^2 	& \Omega 	& 1 \\
1 			& \Omega^2 	& \Omega \\
\Omega		& 1 			& \Omega^2  
\end{bmatrix}
\end{align*}
and 
\begin{align*}
\abcT_{\Omega^*} = \frac{1}{(\Omega^*-1)(\Omega^* - \Omega)}
\begin{bmatrix}
\Omega^{*2} 	& \Omega^* 	& 1 \\
1 			& \Omega^{*2} 	& \Omega^* \\
\Omega^*		& 1 			& \Omega^{*2}  
\end{bmatrix}~.
\end{align*}
From Eq.~\eqref{eq:DiagonalizableCorrelationAmplitudes}, we have:
\begin{align*}
\Braket{\abcT_\Omega ^{\coperator(\Alphabet)}} &= \frac{\Omega^2}{(\Omega-1)(\Omega-\Omega^*)}~, \\
\Braket{\abcT_\Omega ^{\aoperator(\Alphabet)}} &= \frac{\Omega}{(\Omega-1)(\Omega-\Omega^*)}~, \\
\Braket{\abcT_{\Omega^*} ^{\coperator(\Alphabet)}} &= \frac{\Omega^{*2}}{(\Omega^*-1)(\Omega^*-\Omega)}~, \\
\intertext{and}
\Braket{\abcT_{\Omega^*} ^{\aoperator(\Alphabet)}} &= \frac{\Omega^*}{(\Omega^*-1)(\Omega^*-\Omega)}~,
\end{align*}
which from Eq.~\eqref{eq:DeltaFnMagnitudes} yields:
\begin{align*}
  \begin{array}{r@{\mskip\thickmuskip}l}
        \Delta_{\Omega} &= 1 
  \end{array} 
   \:\:\: {\rm {and}} \:\:\:
  \begin{array}{r@{\mskip\thickmuskip}l}
\Delta_{\Omega^*} &= 0~.
  \end{array}
\end{align*}
Then, using Eq.~\eqref{eq:DiscreteDP}, the DP's discrete part becomes:
\begin{align*}
\DP_{{\mathrm {B}}}(\ell) & =
\sum_{k = -\infty}^{\infty}  \delta(\ell - \tfrac{1}{3} + k)~,
\end{align*}
as it ought to be for 3C$^+$. 

\subsection{Random Growth and Deformation Faults in Layered 3C and 2H CPSs: The RGDF Process}
\label{GrowthDeformationStacking}

As a simple model of faulting in CPSs, combined random growth and deformation
faults are often assumed if the faulting probabilities are believed to be
small. However, until now there has not been an analytical expression available
for the DP for all values of the faulting parameters, and we derive such an
expression here.

\begin{figure}
\begin{center}
\includegraphics[width=0.47\textwidth]{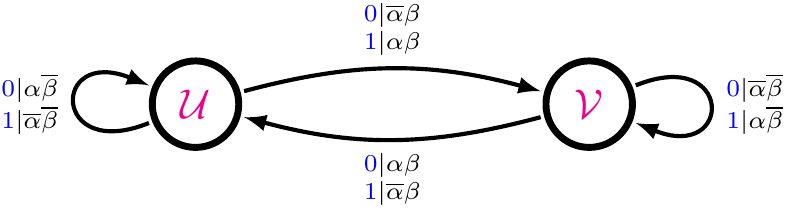}
\end{center}
\caption{\MachineHagg\ for the RGDF Process as proposed by
  Estevez~\etal~(\citeyear{Este08a}). This two-state machine has two
  parameters, $\alpha \in [0, 1]$ and $\beta \in [0,1]$, the probability of
  deformation and growth faults in CPSs, respectively. (From
  Riechers~\etal~(\citeyear{Riec14b}), used with permission.)
  }
\label{fig:EstevezHagg}
\end{figure}

The HMM for the \emph{Random Growth and Deformation Faults} (RGDF) process was
first proposed by Estevez-Rams \etal\ (\citeyear{Este08a}) and the
\MachineHagg\ is shown in Fig.~\ref{fig:EstevezHagg}. The process has two
parameters, $\alpha \in [0, 1]$ and $\beta \in [0,1]$, that (at least for small
values) are interpreted as the probability of deformation and growth faults,
respectively. The stacking process, however, is described best on its own
terms---in terms of the HMM, which captures the causal architecture of the
stacking for all parameter values.

It is instructive to consider limiting values of $\alpha$ and $\beta$.  For
$\alpha = \beta = 0$, the stacking structure is simply 3C. The machine splits
into two distinct machines: each machine has one state with a single self-state
transition, corresponding to the 3C$^+$ stacking structure and the other to
3C$^-$ stacking structure. The 2H stacking structure occurs when $\alpha =
\bar{\beta} = 0$. Typically growth faults are introduced as $\beta$ strays from
these limiting values, and deformation faults appear when $\alpha$ becomes
small but nonvanishing. When $\alpha = 1/2$, the stacking becomes
completely random, regardless of the value of $\beta$. This is summarized in
Table~\ref{Table:RandomDefGrowthFault}.

\begin{table}
\caption{\label{Table:RandomDefGrowthFault} The limiting material structures
  for the RGDF Process. Key: GF - growth fault; DF - deformation fault; Ran -
  completely random stacking.
  }
\begin{ruledtabular}
\begin{tabular}{l|ccccc}
                       & $\beta = 0$  & $\beta \approx 0$  & $\beta = \bar{\beta} = 1/2 $  & $\bar{\beta} \approx 0$ &  $\bar{\beta} = 0$ \\  \hline
    $\alpha = 0$   &   3C    &  3C/GF   &   Ran     &  2H/GF   &  2H \\
    $\alpha \approx 0$   &   3C/DF    &  3C/DF,GF  &   Ran     &  2H/DF,GF    & 2H/DF \\
    $\alpha = \frac{1}{2} $   &   Ran    &  Ran   &   Ran   &   Ran   & Ran \\
\end{tabular}
\end{ruledtabular}
\end{table}

The RGDF \MachineHagg's TMs are:
\begin{align*}
  \begin{array}{r@{\mskip\thickmuskip}l}
     \HaggTzero & = 
         \begin{bmatrix}
         \alpha \overline{\beta} & \overline{\alpha} \beta \\
         \alpha \beta & \overline{\alpha} \overline{\beta}
         \end{bmatrix} 
  \end{array} 
{\rm {and}}
  \begin{array}{r@{\mskip\thickmuskip}l}
      \HaggTone & = 
          \begin{bmatrix}
          \overline{\alpha} \overline{\beta} & \alpha \beta \\
          \overline{\alpha} \beta & \alpha \overline{\beta}
    \end{bmatrix} . 
  \end{array}
\end{align*}
The \MachineHagg\ is
nonmixing only for the parameter settings $\beta = 1$ and $\alpha \in \{0,1\}$,
giving rise to 2H crystal structure.

From the \Hagg-machine, we obtain the corresponding TMs of the $ABC$-machine for 
$\alpha, \beta \in (0, 1)$:~\cite{Riec14b}
\begin{align}
\abcTA & = 
    \begin{bmatrix} \nonumber
       0 & 0 & 0 & 0 & 0 & 0 \\
       \alpha \overline{\beta} & 0 & 0 & \overline{\alpha} \beta & 0 & 0 \\
       \overline{\alpha} \overline{\beta} & 0 & 0 & \alpha \beta & 0 & 0 \\
       0 & 0 & 0 & 0 & 0 & 0 \\
       \alpha \beta & 0 & 0 & \overline{\alpha} \overline{\beta} & 0 & 0 \\
       \overline{\alpha} \beta & 0 & 0 & \alpha \overline{\beta} & 0 & 0 \\
    \end{bmatrix} , \\ \vspace{.05in}
\abcTB & = 
    \begin{bmatrix} \nonumber
        0 & \overline{\alpha} \overline{\beta} & 0 & 0 & \alpha \beta & 0 \\
        0 & 0 & 0 & 0 & 0 & 0 \\
        0 & \alpha \overline{\beta} & 0 & 0 & \overline{\alpha} \beta & 0 \\
        0 & \overline{\alpha} \beta & 0 & 0 & \alpha \overline{\beta} & 0 \\
        0 & 0 & 0 & 0 & 0 & 0 \\
        0 & \alpha \beta & 0 & 0 & \overline{\alpha} \overline{\beta} & 0 \\
    \end{bmatrix} ,
\intertext{and}
\abcTC & = 
    \begin{bmatrix} \nonumber
        0 & 0 & \alpha \overline{\beta} & 0 & 0 & \overline{\alpha} \beta \\
        0 & 0 & \overline{\alpha} \overline{\beta} & 0 & 0 & \alpha \beta \\
        0 & 0 & 0 & 0 & 0 & 0 \\
        0 & 0 & \alpha \beta & 0 & 0 & \overline{\alpha} \overline{\beta} \\
        0 & 0 & \overline{\alpha} \beta & 0 & 0 & \alpha \overline{\beta} \\
        0 & 0 & 0 & 0 & 0 & 0 \\
    \end{bmatrix} , 
\end{align}
and the orientation-agnostic state-to-state TM:
\begin{align}
\abcT & = \abcTA + \abcTB + \abcTC .   \nonumber
\end{align}
Explicitly, we have: 
\begin{align}
\abcT & = 
    \begin{bmatrix}   \nonumber
        0 & \overline{\alpha} \overline{\beta} & \alpha \overline{\beta} & 0 & \alpha \beta & \overline{\alpha} \beta \\
        \alpha \overline{\beta} & 0 & \overline{\alpha} \overline{\beta} & \overline{\alpha} \beta & 0 & \alpha \beta \\
        \overline{\alpha} \overline{\beta} & \alpha \overline{\beta} & 0 & \alpha \beta & \overline{\alpha} \beta & 0 \\
        0 & \overline{\alpha} \beta & \alpha \beta & 0 & \alpha \overline{\beta} & \overline{\alpha} \overline{\beta} \\
        \alpha \beta & 0 & \overline{\alpha} \beta & \overline{\alpha} \overline{\beta} & 0 & \alpha \overline{\beta} \\
        \overline{\alpha} \beta & \alpha \beta & 0 & \alpha \overline{\beta} & \overline{\alpha} \overline{\beta} & 0 \\
    \end{bmatrix} .
\end{align}
$\abcT$'s eigenvalues satisfy det$(\abcT - \lambda \IdentMat) = 0$, 
from which we obtain the eigenvalues:~\cite{Riec14b}
\begin{align}
\Lambda_{\abcT} 
\label{eq:EstevezEigsWSigma}
& = \left\{ 1, \, 1 - 2\beta, \, -\tfrac{1}{2} (1 - \beta) \pm \tfrac{1}{2} \sqrt{\sigma}  \right\} , 
\end{align}
with 
\begin{align}
\label{eq:ExpansionOfSigma}
\sigma &\equiv
4\beta^2 - 3\overline{\beta}^2 + 12 \alpha \overline{\alpha}(\overline{\beta} - \beta) \\
& = -3 + 12\alpha + 6\beta - 12\alpha^2 + \beta^2 - 24\alpha \beta + 24\alpha^2 \beta .
\end{align}

Except for measure-zero submanifolds along which the eigenvalues become extra
degenerate, throughout the parameter range the eigenvalues' algebraic
multiplicities are: $a_1 = 1$, $a_{1-2\beta} = 1$, $a_{- \tfrac{1}{2} (1 -
\beta + \sqrt{\sigma})} = 2$, and $a_{- \tfrac{1}{2} (1 - \beta -
\sqrt{\sigma})} = 2$. Moreover, the \emph{index} of all eigenvalues is 1 except
along $\sigma = 0$. Hence, due to their qualitative difference, we treat the
cases of $\sigma = 0$ and $\sigma \neq 0$ separately.

\begin{figure*}
\begin{center}
\resizebox{!}{6.0in}{\includegraphics{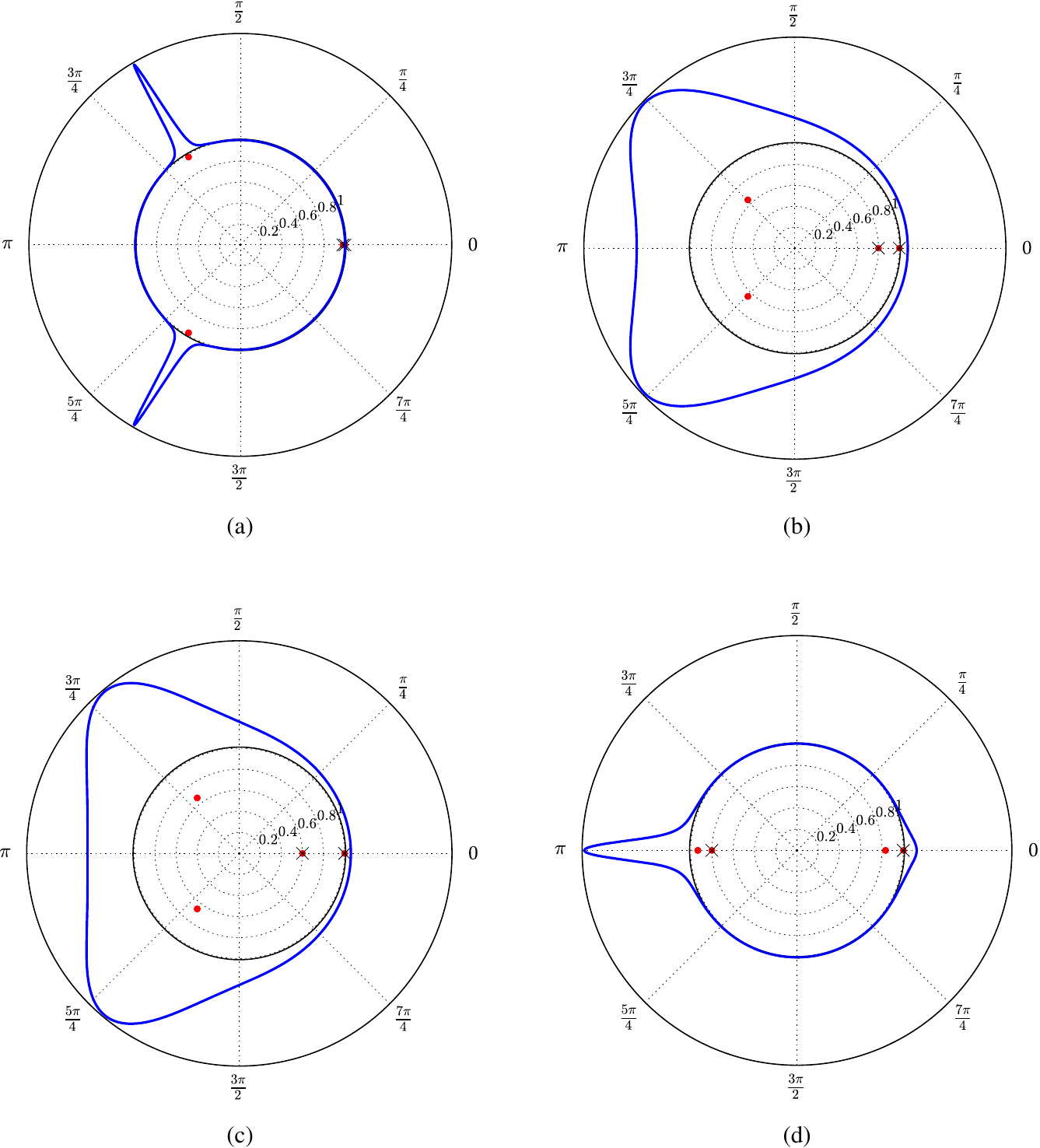}}
\end{center}     
\caption{\Csg s showing the DP and eigenvalues for the RGDF Process:
  (a) $\alpha = 0.01$, $\beta = 0.01$; (b) $\alpha = 0.2$, $\beta = 0.1$; (c)
  $\alpha = 0.1$, $\beta = 0.2$; and (d) $\alpha = 0.01$, $\beta = 0.9$. Note
  how the eigenvalues organize the DP: as the eigenvalues approach the unit
  circle, the DP becomes enhanced. Also, note that nowhere is there enhanced
  scattering without an underlying eigenvalue of the TM driving it.
  }
\label{Fig:RGDFCoronalSpectroGrams} 
\end{figure*}

\subsubsection{$\sigma=0$:}

Riechers \etal~\cite{Riec14b} found that:
\begin{align*}
\Braket{\abcT_1 ^{\coperator(\Alphabet)}} = \Braket{\abcT_1 ^{\aoperator(\Alphabet)}}  &= \tfrac{1}{3}~, \\ 
\Braket{\abcT_{1-2\beta} ^{\coperator(\Alphabet)}} = \Braket{\abcT_{1-2\beta} ^{\aoperator(\Alphabet)}} &= 0~, \\ 
\Braket{\abcT_{-\overline{\beta}/2} ^{\coperator(\Alphabet)}} = 
\Braket{\abcT_{-\overline{\beta}/2} ^{\aoperator(\Alphabet)}} &= \tfrac{1}{6}~, \\  
\intertext{and} 
\Braket{\abcT_{-\overline{\beta}/2, 1} ^{\coperator(\Alphabet)}} = 
\Braket{\abcT_{-\overline{\beta}/2, 1} ^{\aoperator(\Alphabet)}} &= 
-\tfrac{1}{12} \beta \overline{\beta}
  ~,
\end{align*}
for the case of $\sigma = 0$.
According to Eq.~\eqref{eq:SymmetricSpectralDiffuseDP}, the DP for $\sigma=0$ is thus:
\begin{align}
\DP_{{\mathrm {D}}}(\ell) 
& = 
- 6 \realpart \left\{ \sum_{\lambda \in \Lambda_{\abcT}} \sum_{m=0}^{\nu_\lambda-1} 
    \frac{\Braket{\abcT_{\lambda, m} ^{\coperator(\Alphabet)}}}{(z-\lambda)^{m+1}}  \right\} \nonumber \\
& = -6 \realpart \left\{  
    \frac{\Braket{\abcT_{1} ^{\coperator(\Alphabet)}}}{z-1}  
    + \frac{\Braket{\abcT_{-\bar{\beta}/2} ^{\coperator(\Alphabet)}}}{z+\bar{\beta}/2} 
    + \frac{\Braket{\abcT_{-\bar{\beta}/2, 1} ^{\coperator(\Alphabet)}}}{\left(z+\bar{\beta}/2 \right)^2}   
    \right\}  \nonumber \\
& = \realpart \left\{  
    -\frac{2}{z-1}  
    - \frac{1}{z+\bar{\beta}/2} 
    + \frac{\beta \bar{\beta}/2}{\left(z+\bar{\beta}/2 \right)^2}   
    \right\}  \nonumber \\
& = 1 - \realpart \left\{  
    \frac{ z + \bar{\beta}^2 / 2}{\left(z+\bar{\beta}/2 \right)^2}   
    \right\}~.     
\end{align}

\subsubsection{$\sigma \neq 0$:}

Riechers \etal~\cite{Riec14b} also found that:
\begin{align*}
\Braket{\abcT_1 ^{\coperator(\Alphabet)}} &= \Braket{\abcT_1 ^{\aoperator(\Alphabet)}}  = \tfrac{1}{3}~, \\ 
\Braket{\abcT_{1-2\beta} ^{\coperator(\Alphabet)}} &= \Braket{\abcT_{1-2\beta} ^{\aoperator(\Alphabet)}} = 0~, \\ 
\Braket{T_{\frac{-\overline{\beta} + \sqrt{\sigma}}{2}}^{\coperator(\Alphabet)}}
  & = \Braket{T_{\frac{-\overline{\beta} + \sqrt{\sigma}}{2}}^{\aoperator(\Alphabet)}}
  \\
  & = - \tfrac{1}{12} \left( 1 - \frac{\beta}{\sqrt{\sigma}} \right)
    \left( \sqrt{\sigma} - \overline{\beta} \right) ~, \\
\intertext{and}     
\Braket{T_{\frac{-\overline{\beta} - \sqrt{\sigma}}{2}}^{\coperator(\Alphabet)}}
  & = \Braket{T_{\frac{-\overline{\beta} - \sqrt{\sigma}}{2}}^{\aoperator(\Alphabet)}}
  \\
  & = \tfrac{1}{12} \left( 1 + \frac{\beta}{\sqrt{\sigma}} \right)
    \left( \sqrt{\sigma} + \overline{\beta} \right)   
  ~,
\end{align*}
for $\sigma \neq 0$.
According to Eq.~\eqref{eq:SymmetricSpectralDiffuseDP}, the DP for $\sigma \neq 0$ is:
\begin{align}
\DP_{{\mathrm {D}}}(\ell) 
& = 
- 6 \realpart \left\{ \sum_{\lambda \in \Lambda_{\abcT}} 
    \frac{\Braket{\abcT_{\lambda} ^{\coperator(\Alphabet)}}}{z-\lambda}  \right\} \nonumber \\
& = -6 \realpart \left\{  
    \frac{\Braket{\abcT_{1} ^{\coperator(\Alphabet)}}}{z-1}  
    + \frac{\Braket{\abcT_{\frac{-\overline{\beta} + \sqrt{\sigma}}{2}} ^{\coperator(\Alphabet)}}}{z-\frac{\sqrt{\sigma} - \overline{\beta} }{2}} 
    + \frac{\Braket{\abcT_{\frac{-\overline{\beta} - \sqrt{\sigma}}{2}} ^{\coperator(\Alphabet)}}}{z+\frac{\sqrt{\sigma} + \overline{\beta}}{2}}   
    \right\}  \nonumber \\
& = 1 + \tfrac{1}{2} \realpart \left\{  
    \frac{ 1 - \beta / \sqrt{\sigma} }{\frac{z}{\sqrt{\sigma} - \bar{\beta}} - \frac{1}{2}}   
    - \frac{ 1 + \beta / \sqrt{\sigma} }{\frac{z}{\sqrt{\sigma} + \bar{\beta}} + \frac{1}{2}}  
    \right\}~.     
\end{align}

Figure~\ref{Fig:RGDFCoronalSpectroGrams} gives several \csg s for various
values of the parameters $\alpha$ and $\beta$. It is instructive to examine the
influence of the TM's eigenvalues on the placement and intensity of the
Bragg-like reflections. In panel (a) there are two strong reflections, one each
at $\omega = 2\pi/3$ and $4\pi/3$, signaling a twinned 3C structure, when the
faulting parameters are set to $\alpha = \beta = 0.01$. Each is accompanied by
an eigenvalue close to the surface of the unit circle. As the disorder is
increased, see panels (b) and (c), TM eigenvalues retreat toward the center of
the unit circle and the two strong reflections become diffuse. However, in the
final panel (d), the faulting parameters ($\alpha = 0.01$, $\beta = 0.9$) are
set such that the material has apparently undergone a phase transition from
prominently 3C stacking structure to prominently 2H stacking structure. Indeed,
the eigenvalues have coalesced through the critical point of $\sigma=0$ (as
$\sigma$ changes from negative to positive) and emerge on the other side of the
phase transition mutually scattered along the real axis and approaching the
edge of the unit circle, giving rise to the 2H-like protrusions in the DP.
This demonstrates again how the eigenvalues orchestrate the placement and
intensity of the Bragg-like peaks.

\subsection{Shockley--Frank Stacking Faults in 6H-SiC: The SFSF Process}
\label{ShockleyFrankStacking}

SiC has been the intense focus of both experimental and theoretical
investigations for some time due to its promise as a material suitable for
next-generation electronic devices. However, it is known that SiC can have many
different stacking configurations---some ordered and some
disordered\cite{Seba94a}---and these different stacking configurations can
profoundly affect material properties. Despite considerable effort to grow
commercial SiC wafers that are purely crystalline---\ie, that have no stacking
defects---reliable techniques have not yet been developed. It is therefore
important to better understand and characterize the nature of the defects in
order to better control them.

Recently, Sun \etal~\cite{Sun12a} reported experiments on 6H-SiC that used a
combination of low temperature photoluminescence and high resolution
transmission electron microscopy (HRTEM). One of the more common crystalline
forms of SiC, the 6H stacking structure is simply the sequence \dots ABCACBA
\dots, or in terms of the \Hagg-notation, \dots 111000 \dots. The most common
stacking fault in 6H-SiC identified by HRTEM can be explained as the result of
one extrinsic Frank stacking fault coupled with one Shockley stacking
fault.~\cite{Hirt68a} Physically, the resultant stacking structure corresponds
to the insertion of an additional SiC ML so that one has instead \dots
$11000\underline{0}111000$ \dots, where the underlined spin is the inserted ML. 

\begin{figure}
\begin{center}
\includegraphics[width=0.47\textwidth]{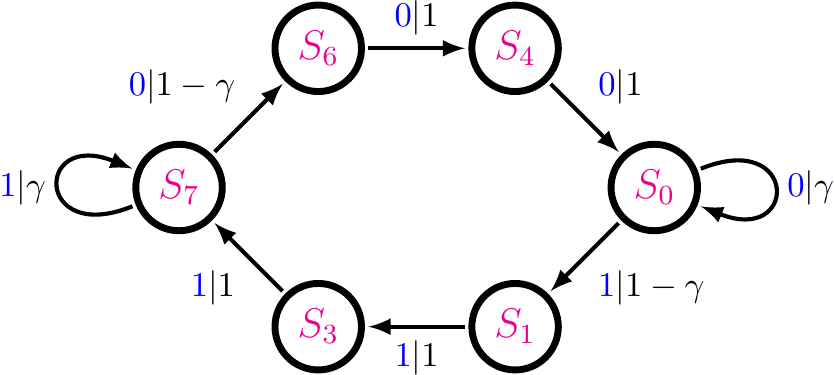}
\end{center}
\caption{\MachineHagg\ for the SFSF Process. There is one faulting parameter
  $\gamma \in [0,1]$ and three SSCs or, equivalently, three CSCs, as this
  machine is also an \eM. The three SSCs are [$\sf{S}_7$], [$\sf{S}_0]$ and
  [$\sf{S}_7 \sf{S}_6 \sf{S}_4 \sf{S}_0 \sf{S}_1 \sf{S}_3$]. The latter we
  recognize as the 6H structure if $\gamma = 0$. For large values of
  $\gamma$---\ie, as $\gamma \to 1$---this process approaches a twinned 3C
  structure, although the faulting is \emph{not} random. The causal-state
  architecture prevents the occurrence of domains of size-three or less. (From
  Riechers~\etal~\cite{Riec14b} Used with permission.)
  }
\label{fig:SunHaggMachine}
\end{figure}

Inspired by these findings, we suggest a simple HMM for the
\emph{Shockley--Frank stacking fault} (SFSF) process that replicates this
structure, and this is shown in Fig.~\ref{fig:SunHaggMachine}. Our motivation
here is largely pedagogical, and certainly more detailed experiments are
required to confidently propose a structure, but this HMM reproduces at least
qualitatively the observed structure. The model has a single parameter $\gamma
\in [0,1]$. As before, it is instructive to consider limiting cases of
$\gamma$. For $\gamma = 0$, we have the pure 6H structure and, for small
$\gamma$, Shockley--Frank defects are introduced into this stacking structure.
As $\gamma \to 1$, the structure transitions into a twinned 3C crystal.
However, unlike the previous example, this twinning is not random. Instead, the
architecture of the machine requires that at least three 0s or 1s must be seen
before there is a possibility of reversing the chirality, \ie, before there is
twinning. These limiting cases are summarized in
Table~\ref{Table:ShockleyFrankFault}.

\begin{table}
\caption{\label{Table:ShockleyFrankFault} Limiting material structures for the
  SFSF Process. Key: SF - Shockley--Frank fault; NGF - nonrandom growth fault.}
\begin{ruledtabular}
\begin{tabular}{cccc}
   $\gamma = 0$  & $\gamma \approx 0$  &  $\bar{\gamma} \approx 0$  &  $\bar{\gamma} = 0$ \\  \hline
    6H   &   6H/SF    &   3C/NGF &    3C    \\
\end{tabular}
\end{ruledtabular}
\end{table}

\begin{figure*}
\begin{center}
\resizebox{!}{6.0in}{\includegraphics{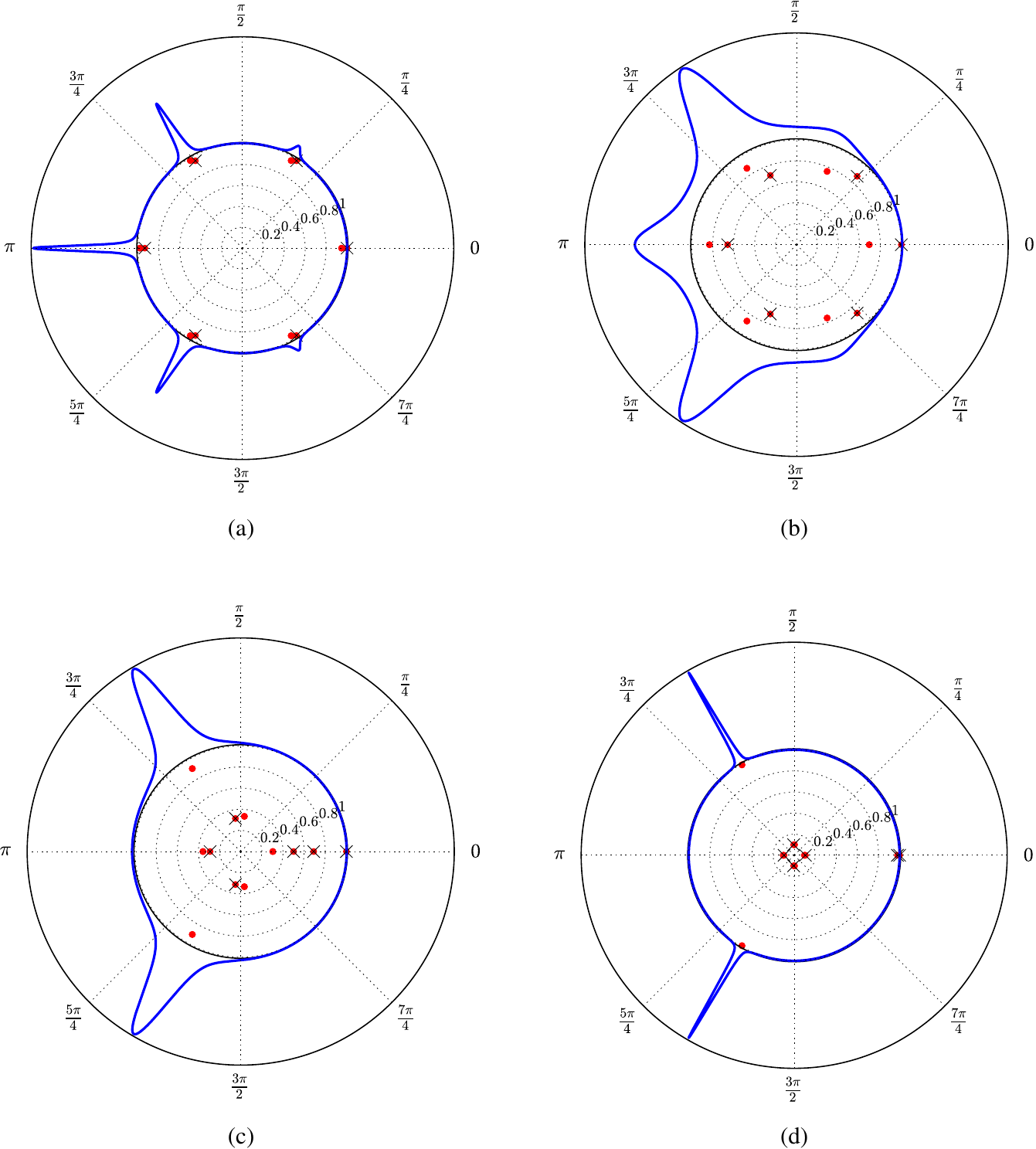}}
\end{center}     
\caption{\Csg s showing the evolution of the DP and its eigenvalues
  for the SFSF Process. (a) $\gamma = 0.1$, (b) $\gamma = 0.5$, (c) $\gamma =
  0.9$, and (d) $\gamma = 0.99$. In (a) the faulting is weak and the DP has the
  six degraded Bragg-like reflections characteristic of the 6H stacking
  structure. In (b), the faulting is more severe, with the concomitant erosion
  of the Bragg-like reflections, especially for $\omega = \pi$. In panel (c)
  the 6H character has been eliminated, and the Bragg-like peaks at $\omega =
  2\pi/3$ and $4\pi/3$ are now associated with a twinned 3C stacking structure.
  In panel (d), the Bragg-like reflections sharpen as the probability of short
  3C sequences stacking sequences decreases.
  } 
\label{Fig:SFSFCoronalSpectroGrams} 
\end{figure*}

For $\gamma \in (0,1)$ the \MachineHagg\ is mixing and we proceed with this
case. By inspection, we write down the two 6-by-6 TMs of the \MachineHagg\ as:
\begin{align}
\HaggTzero & = 
  \begin{bmatrix} \nonumber 
  \gamma & 0 & 0 & 0 & 0 & 0 \\
  0 & 0 & 0 & 0 & 0 & 0 \\
  0 & 0 & 0 & 0 & 0 & 0 \\ 
  0 & 0 & 0 & 0 & \overline{\gamma} & 0 \\ 
  0 & 0 & 0 & 0 & 0 & 1 \\ 
  1 & 0 & 0 & 0 & 0 & 0 \\ 
  \end{bmatrix}
\intertext{and:}
\HaggTone & = 
  \begin{bmatrix} \nonumber 
  0 & \overline{\gamma} & 0 & 0 & 0 & 0 \\
  0 & 0 & 1 & 0 & 0 & 0 \\
  0 & 0 & 0 & 1 & 0 & 0 \\ 
  0 & 0 & 0 & \gamma & 0 & 0 \\ 
  0 & 0 & 0 & 0 & 0 & 0 \\ 
  0 & 0 & 0 & 0 & 0 & 0 \\ 
  \end{bmatrix} ,
\end{align}
where the states are ordered $\sf{S}_0$, $\sf{S}_1$, $\sf{S}_3$, $\sf{S}_7$,
$\sf{S}_6$, and $\sf{S}_4$. The internal state TM is their sum:
\begin{align*}
\HaggT & = 
  \begin{bmatrix} \nonumber 
  \gamma & \overline{\gamma} & 0 & 0 & 0 & 0 \\
  0 & 0 & 1 & 0 & 0 & 0 \\
  0 & 0 & 0 & 1 & 0 & 0 \\ 
  0 & 0 & 0 & \gamma & \overline{\gamma} & 0 \\ 
  0 & 0 & 0 & 0 & 0 & 1 \\ 
  1 & 0 & 0 & 0 & 0 & 0 \\ 
  \end{bmatrix} .
\end{align*}
Since the six-state \MachineHagg\ generates an ($3 \times 6 = $) eighteen-state \MachineABC, we do not explicitly write out its TMs. Nevertheless, it is straightforward to expand the \MachineHagg\ to the \MachineABC\ via the rote expansion method.~\cite{Riec14b} It is also straightforward to apply Eq.~\eqref{eq:MixingDiffuseDP} to obtain the DP as a function of the faulting parameter $\gamma$. To use Eq.~\eqref{eq:MixingDiffuseDP}, note that the stationary distribution over the \MachineABC\ can be obtained from 
Eq.~\eqref{eq:ABCPiFromHaggPi} with:
\begin{align*}
\bra{\Dist_{\textrm{H}}} &= 
  \tfrac{1}{6 - 4 \gamma} 
  \begin{bmatrix}  1 & \overline{\gamma} & \overline{\gamma} & 1 & \overline{\gamma} & \overline{\gamma}  \end{bmatrix} 
\end{align*}
as the stationary distribution over the \MachineHagg. 

The eigenvalues of the \Hagg\ TM can be obtained as the solutions of
$\det(\HaggT - \lambda \IdentMat) = (\lambda - \gamma)^2 \lambda^4 -
\overline{\gamma}^2 = 0$.  These include $1$, $-\tfrac{1}{2} \overline{\gamma}
\pm \sqrt{\gamma^2 + 2\gamma - 3}$, and three other eigenvalues involving cube
roots.

The eigenvalues of the $ABC$ TM are obtained similarly as the solutions of
$\det(\abcT - \lambda \IdentMat) = 0$. Note that $\Lambda_{\abcT}$ inherits
$\Lambda_{\HaggT}$ as the backbone for its more complex structure, just as
$\Lambda_{\HaggT} \subseteq \Lambda_{\abcT}$ for all of our previous examples.
The eigenvalues in $\Lambda_{\abcT}$ are, of course, those most directly
responsible for the structure of the CFs. Since the \MachineABC\ has eighteen
states, there are eighteen eigenvalues contributing to the behavior of the DP;
although several eigenvalues are degenerate. Hence, the SFSF Process is capable
of a richer DP than the previous two examples. 

The \csg s for the SFSF Process are shown for several example values of
$\gamma$ in Fig.~\ref{Fig:SFSFCoronalSpectroGrams}. Over the range of $\gamma$
values the stacking structure changes from a nearly perfect 6H crystal through
a disordered phase finally becoming a twinned 3C structure. Most notable in
Fig.~\ref{Fig:SFSFCoronalSpectroGrams} is how the eigenvalues of the total TM
dictate the placement of the Bragg-like reflections. Phrased alternatively, the
Bragg-like reflections appear to literally track the movement of the
eigenvalues as they evolve during transformation.

\section{Conclusions}
\label{Conclusions}

We showed how the DP for an arbitrary HMM stacking process is calculated either
analytically or to a high degree of numerical certainty directly without
restriction to finite Markov order and without needing finite samples of the
stacking sequence. Along the way, we uncovered a remarkably simple relationship
between the DP and the HMM. The former is given by straightforward, standard
matrix manipulations of the latter. Critically, in the case of an infinite
number of MLs, this relationship does not involve powers of the TM.

The connection yields important insights. (i) The number of Bragg and
Bragg-like reflections in the DP is limited by the size of the TMs that define
the HMM. Thus, knowing only the number of machine states reveals the maximum
possible number of Bragg and Bragg-like reflections. (ii) As a corollary, given
a DP, the number of Bragg and Bragg-like reflections puts a minimum on the
number of HMM states. For the problem of inferring the HMM from experimental
DPs, this gives powerful clues about the HMM (and so internal mechanism)
architecture. (iii) The eigenvalues \emph{within} the unit circle organize the
diffuse Bragg-like reflections. Only TM eigenvalues \emph{on} the unit circle
correspond to those $\ell$-values that potentially can result in true Bragg
peaks. (iv) The expansion of the \MachineHagg\ into the \MachineABC, necessary
for the appropriate matrix manipulations, showed that there are two kinds of
machines and, hence, two kinds of stacking process important in CPSs: mixing
and nonmixing processes. In addition to the calculational shortcuts given by
the former, mixing machines ensure that there are no true Bragg reflections at
integer-$\ell$. (v) Conversely, the presence of Bragg peaks at integer-$\ell$
is an unmistakable sign that a stacking process is nonmixing. Again, this puts
important constraints on the HMM state architecture, useful for the problem of
inverting the DP to find the HMM. (vi) For mixing processes, the ML
probabilities must all be one-third, \ie, $\Pr(A) = \Pr(B) = \Pr(C) = 1/3$.

New in the theory is the introduction of \csg s, a convenient way to visualize
the interplay between a frequency-domain functional of a process and the
eigenvalues of the process's TMs. In our case, the frequency-domain functional
was the DP: the power spectrum of the sequence of ML structure factors. In each
of the examples, the movement of the eigenvalues (as the HMM parameters change)
were echoed by movement of their `shadow'---the Bragg-like peaks in the power
spectrum. While this technique was explored in the context of DPs from layered
materials, this visualization tool is by no means confined to DPs or layered
materials. We suspect that in other areas where power spectra and HMMs are
studied, this technique will become a useful analysis tool.

There are several important research directions to follow in further refining
and extending the theory and developing applications. (i) While specialized to
CPSs here, the basic techniques extend to other stacking geometries and other
materials, including the gamut of technologically cutting-edge heterostructures
of stacked 2D materials. (ii) With the ability to analytically calculate DPs
and CFs~\cite{Riec14b} from arbitrary HMMs, the number of physical and
information- and computation-theoretic quantities amenable to such a treatment
continues to expand. Statistical complexity, the Shannon entropy rate, and
memory length have long been calculable from the
\eM,~\cite{Crut89a,Varn07a,Varn13a,Varn13b} but recently the excess entropy,
transient information, and synchronization time have also been shown to be
exactly calculable from the \eM.~\cite{Crut13a,Riec14a} This portends well that
additional quantities, especially those of physical import such as band
structure in chaotic crystals, may also be treatable with exact methods. (iii)
Improved calculational techniques raise the possibility of improved inference
methods, so that more kinds of stacking process may be discovered from DPs. An
important research direction then is to incorporate these improved methods into
more flexible, more sensitive inference algorithms.

Finally, the spectral methods pursued here increase the tools available to
chaotic crystallography for the discovery, description, and categorization of
both ordered and disordered (chaotic) crystals. With these tools in hand, we
will more readily identify key features of the hidden structures responsible
for novel physical properties of materials.

\begin{acknowledgments}
The authors thank the Santa Fe Institute for its hospitality during visits.
JPC is an External Faculty member there. This material is based upon work
supported by, or in part by, the U.~S.~Army Research Laboratory and the 
U.~S.~Army Research Office under contract number W911NF-13-1-0390.
\end{acknowledgments}

\appendix
\section{\Hagg-to-$ABC$ Machine Translation}

If $\NstatesHagg$ is the number
of states in the \MachineHagg\ and $\Nstates$ is the number of states in the \MachineABC, 
then $\Nstates = 3 \NstatesHagg$ for mixing \MachineHagg s.
Let the $i^{\rm {th}}$ state of the \MachineHagg\ split into 
the ${\left( 3i-2 \right) }^{\rm {th}}$ through the ${\left( 3i \right) }^{\rm
{th}}$ states of the 
corresponding \MachineABC. Then, each labeled-edge transition from 
the $i^{\rm {th}}$ to the $j^{\rm {th}}$ states of the \MachineHagg\ maps into a 
3-by-3 submatrix for each of the three labeled TMs of the 
\MachineABC\ as:
\begin{align*}
\left\{\TransMatHagg_{ij}^{[0]} \right\} 
\xRightarrow{\text{\Hagg\ to } ABC} 
\left\{ \TransMatABC_{3i-1, 3j-2}^{[A]}, \text{ } \TransMatABC_{3i, 3j-1}^{[B]}, \text{ } \TransMatABC_{3i-2, 3j}^{[C]} \right\}
\end{align*}
and
\begin{align*}
\left\{\TransMatHagg_{ij}^{[1]} \right\} 
\xRightarrow{\text{\Hagg\ to } ABC} 
\left\{ \TransMatABC_{3i, 3j-2}^{[A]}, \text{ } \TransMatABC_{3i-2, 3j-1}^{[B]}, \text{ } \TransMatABC_{3i-1, 3j}^{[C]} \right\} .
\end{align*}

For nonmixing \MachineHagg s, the above algorithm creates three disconnected
\MachineABC s, of which only one need be retained.

Furthermore, for mixing \MachineHagg s, the probability from the stationary
distribution over their states maps to a triplet of probabilities for the
stationary distribution over the \MachineABC\ states: 
\begin{align}
\left\{ {p_i^{\text{H}}} \right\} 
\xRightarrow{\text{\Hagg\ to } ABC} 
\left\{ 3p_{3i-2}, \text{ } 3p_{3i-1}, \text{ } 3p_{3i} \right\} ~.
\label{eq:ABCPiFromHaggPi}
\end{align}

A thorough exposition of these procedures is given by Riechers
\etal.~\cite{Riec14b}

\bibliography{iucr}

\begin{thebibliography}{68}%
\makeatletter
\providecommand \@ifxundefined [1]{%
 \@ifx{#1\undefined}
}%
\providecommand \@ifnum [1]{%
 \ifnum #1\expandafter \@firstoftwo
 \else \expandafter \@secondoftwo
 \fi
}%
\providecommand \@ifx [1]{%
 \ifx #1\expandafter \@firstoftwo
 \else \expandafter \@secondoftwo
 \fi
}%
\providecommand \natexlab [1]{#1}%
\providecommand \enquote  [1]{``#1''}%
\providecommand \bibnamefont  [1]{#1}%
\providecommand \bibfnamefont [1]{#1}%
\providecommand \citenamefont [1]{#1}%
\providecommand \href@noop [0]{\@secondoftwo}%
\providecommand \href [0]{\begingroup \@sanitize@url \@href}%
\providecommand \@href[1]{\@@startlink{#1}\@@href}%
\providecommand \@@href[1]{\endgroup#1\@@endlink}%
\providecommand \@sanitize@url [0]{\catcode `\\12\catcode `\$12\catcode
  `\&12\catcode `\#12\catcode `\^12\catcode `\_12\catcode `\%12\relax}%
\providecommand \@@startlink[1]{}%
\providecommand \@@endlink[0]{}%
\providecommand \url  [0]{\begingroup\@sanitize@url \@url }%
\providecommand \@url [1]{\endgroup\@href {#1}{\urlprefix }}%
\providecommand \urlprefix  [0]{URL }%
\providecommand \Eprint [0]{\href }%
\providecommand \doibase [0]{http://dx.doi.org/}%
\providecommand \selectlanguage [0]{\@gobble}%
\providecommand \bibinfo  [0]{\@secondoftwo}%
\providecommand \bibfield  [0]{\@secondoftwo}%
\providecommand \translation [1]{[#1]}%
\providecommand \BibitemOpen [0]{}%
\providecommand \bibitemStop [0]{}%
\providecommand \bibitemNoStop [0]{.\EOS\space}%
\providecommand \EOS [0]{\spacefactor3000\relax}%
\providecommand \BibitemShut  [1]{\csname bibitem#1\endcsname}%
\let\auto@bib@innerbib\@empty
\bibitem [{\citenamefont {Hirth}\ and\ \citenamefont {Lothe}(1968)}]{Hirt68a}%
  \BibitemOpen
  \bibfield  {author} {\bibinfo {author} {\bibfnamefont {J.~P.}\ \bibnamefont
  {Hirth}}\ and\ \bibinfo {author} {\bibfnamefont {J.}~\bibnamefont {Lothe}},\
  }\href@noop {} {\emph {\bibinfo {title} {Theory of Dislocations}}},\ \bibinfo
  {edition} {2nd}\ ed.\ (\bibinfo  {publisher} {McGraw-Hill},\ \bibinfo
  {address} {New York},\ \bibinfo {year} {1968})\BibitemShut {NoStop}%
\bibitem [{\citenamefont {Caroff}\ \emph {et~al.}(2011)\citenamefont {Caroff},
  \citenamefont {Bolinsson},\ and\ \citenamefont {Johansson}}]{Caro11a}%
  \BibitemOpen
  \bibfield  {author} {\bibinfo {author} {\bibfnamefont {P.}~\bibnamefont
  {Caroff}}, \bibinfo {author} {\bibfnamefont {J.}~\bibnamefont {Bolinsson}}, \
  and\ \bibinfo {author} {\bibfnamefont {J.}~\bibnamefont {Johansson}},\
  }\href@noop {} {\bibfield  {journal} {\bibinfo  {journal} {IEEE Journal of
  Selected Topics in Quantum Electronics}\ }\textbf {\bibinfo {volume} {17}},\
  \bibinfo {pages} {829} (\bibinfo {year} {2011})}\BibitemShut {NoStop}%
\bibitem [{\citenamefont {Geim}\ and\ \citenamefont
  {Grigorieva}(2013)}]{Geim13a}%
  \BibitemOpen
  \bibfield  {author} {\bibinfo {author} {\bibfnamefont {A.~K.}\ \bibnamefont
  {Geim}}\ and\ \bibinfo {author} {\bibfnamefont {I.~V.}\ \bibnamefont
  {Grigorieva}},\ }\href@noop {} {\bibfield  {journal} {\bibinfo  {journal}
  {Nature}\ }\textbf {\bibinfo {volume} {499}},\ \bibinfo {pages} {419}
  (\bibinfo {year} {2013})}\BibitemShut {NoStop}%
\bibitem [{\citenamefont {Pan}\ \emph {et~al.}(2009)\citenamefont {Pan},
  \citenamefont {Wang}, \citenamefont {Zhao}, \citenamefont {Wu}, \citenamefont
  {Zhang}, \citenamefont {Wang},\ and\ \citenamefont {Jiao}}]{Pan09a}%
  \BibitemOpen
  \bibfield  {author} {\bibinfo {author} {\bibfnamefont {D.}~\bibnamefont
  {Pan}}, \bibinfo {author} {\bibfnamefont {S.}~\bibnamefont {Wang}}, \bibinfo
  {author} {\bibfnamefont {B.}~\bibnamefont {Zhao}}, \bibinfo {author}
  {\bibfnamefont {M.}~\bibnamefont {Wu}}, \bibinfo {author} {\bibfnamefont
  {H.}~\bibnamefont {Zhang}}, \bibinfo {author} {\bibfnamefont
  {Y.}~\bibnamefont {Wang}}, \ and\ \bibinfo {author} {\bibfnamefont
  {Z.}~\bibnamefont {Jiao}},\ }\href {\doibase 10.1021/cm900395k} {\bibfield
  {journal} {\bibinfo  {journal} {Chemistry of Materials}\ }\textbf {\bibinfo
  {volume} {21}},\ \bibinfo {pages} {3136} (\bibinfo {year}
  {2009})}\BibitemShut {NoStop}%
\bibitem [{\citenamefont {Seebauer}\ and\ \citenamefont {Noh}(2010)}]{Seeb10a}%
  \BibitemOpen
  \bibfield  {author} {\bibinfo {author} {\bibfnamefont {E.~G.}\ \bibnamefont
  {Seebauer}}\ and\ \bibinfo {author} {\bibfnamefont {K.~W.}\ \bibnamefont
  {Noh}},\ }\href@noop {} {\bibfield  {journal} {\bibinfo  {journal} {Mater.\
  Sci.\ Eng.\ R}\ }\textbf {\bibinfo {volume} {70}},\ \bibinfo {pages} {151}
  (\bibinfo {year} {2010})}\BibitemShut {NoStop}%
\bibitem [{\citenamefont {{A. L. Mackay}}(1986)}]{Mack86a}%
  \BibitemOpen
  \bibfield  {author} {\bibinfo {author} {\bibnamefont {{A. L. Mackay}}},\
  }\href@noop {} {\bibfield  {journal} {\bibinfo  {journal} {Computers \&
  Mathematics with Applications}\ }\textbf {\bibinfo {volume} {B12}},\ \bibinfo
  {pages} {21} (\bibinfo {year} {1986})}\BibitemShut {NoStop}%
\bibitem [{\citenamefont {Cartwright}\ and\ \citenamefont
  {Mackay}(2012)}]{Cart12a}%
  \BibitemOpen
  \bibfield  {author} {\bibinfo {author} {\bibfnamefont {J.~H.~E.}\
  \bibnamefont {Cartwright}}\ and\ \bibinfo {author} {\bibfnamefont {A.~L.}\
  \bibnamefont {Mackay}},\ }\href@noop {} {\bibfield  {journal} {\bibinfo
  {journal} {Phil.\ Trans.\ R.\ Soc.\ A}\ }\textbf {\bibinfo {volume} {370}},\
  \bibinfo {pages} {2807} (\bibinfo {year} {2012})}\BibitemShut {NoStop}%
\bibitem [{\citenamefont {Varn}\ and\ \citenamefont
  {Crutchfield}(2014)}]{Varn14a}%
  \BibitemOpen
  \bibfield  {author} {\bibinfo {author} {\bibfnamefont {D.~P.}\ \bibnamefont
  {Varn}}\ and\ \bibinfo {author} {\bibfnamefont {J.~P.}\ \bibnamefont
  {Crutchfield}},\ }\href@noop {} {\bibfield  {journal} {\bibinfo  {journal}
  {Santa Fe Institute Working Paper 14-09-036}\ } (\bibinfo {year} {2014})},\
  \Eprint {http://arxiv.org/abs/1409.5930} {arXiv:1409.5930
  [cond-mat.stat-mech]} \BibitemShut {NoStop}%
\bibitem [{\citenamefont {Crutchfield}(2012)}]{Crut12a}%
  \BibitemOpen
  \bibfield  {author} {\bibinfo {author} {\bibfnamefont {J.~P.}\ \bibnamefont
  {Crutchfield}},\ }\href@noop {} {\bibfield  {journal} {\bibinfo  {journal}
  {Nature Physics}\ }\textbf {\bibinfo {volume} {8}},\ \bibinfo {pages} {17}
  (\bibinfo {year} {2012})}\BibitemShut {NoStop}%
\bibitem [{\citenamefont {Shalizi}\ and\ \citenamefont
  {Crutchfield}(2001)}]{Shal01b}%
  \BibitemOpen
  \bibfield  {author} {\bibinfo {author} {\bibfnamefont {C.~R.}\ \bibnamefont
  {Shalizi}}\ and\ \bibinfo {author} {\bibfnamefont {J.~P.}\ \bibnamefont
  {Crutchfield}},\ }\href@noop {} {\bibfield  {journal} {\bibinfo  {journal}
  {J.\ Stat.\ Phys.}\ }\textbf {\bibinfo {volume} {104}},\ \bibinfo {pages}
  {817} (\bibinfo {year} {2001})}\BibitemShut {NoStop}%
\bibitem [{\citenamefont {Cover}\ and\ \citenamefont {Thomas}(2006)}]{Cove06a}%
  \BibitemOpen
  \bibfield  {author} {\bibinfo {author} {\bibfnamefont {T.~M.}\ \bibnamefont
  {Cover}}\ and\ \bibinfo {author} {\bibfnamefont {J.~A.}\ \bibnamefont
  {Thomas}},\ }\href@noop {} {\emph {\bibinfo {title} {Elements of Information
  Theory}}},\ \bibinfo {edition} {2nd}\ ed.\ (\bibinfo  {publisher} {John Wiley
  \& Sons},\ \bibinfo {address} {Hoboken},\ \bibinfo {year} {2006})\BibitemShut
  {NoStop}%
\bibitem [{\citenamefont {Paz}(1971)}]{Paz71a}%
  \BibitemOpen
  \bibfield  {author} {\bibinfo {author} {\bibfnamefont {A.}~\bibnamefont
  {Paz}},\ }\href@noop {} {\emph {\bibinfo {title} {Introduction to
  Probabilistic Automata}}}\ (\bibinfo  {publisher} {Academic Press},\ \bibinfo
  {address} {New York},\ \bibinfo {year} {1971})\BibitemShut {NoStop}%
\bibitem [{\citenamefont {Hopcroft}\ and\ \citenamefont
  {Ullman}(1979)}]{Hopc79a}%
  \BibitemOpen
  \bibfield  {author} {\bibinfo {author} {\bibfnamefont {J.~E.}\ \bibnamefont
  {Hopcroft}}\ and\ \bibinfo {author} {\bibfnamefont {J.~D.}\ \bibnamefont
  {Ullman}},\ }\href@noop {} {\emph {\bibinfo {title} {Introduction to Automata
  Theory, Languages, and Computation}}}\ (\bibinfo  {publisher}
  {Addison-Wesley},\ \bibinfo {address} {Reading},\ \bibinfo {year}
  {1979})\BibitemShut {NoStop}%
\bibitem [{\citenamefont {Strogatz}(2001)}]{Stro01a}%
  \BibitemOpen
  \bibfield  {author} {\bibinfo {author} {\bibfnamefont {S.~H.}\ \bibnamefont
  {Strogatz}},\ }\href@noop {} {\emph {\bibinfo {title} {Nonlinear Dynamics and
  Chaos: With Applications to Physics, Biology, Chemistry, and Engineering}}}\
  (\bibinfo  {publisher} {Westview Press},\ \bibinfo {year} {2001})\BibitemShut
  {NoStop}%
\bibitem [{\citenamefont {Feldman}(2012)}]{Feld12a}%
  \BibitemOpen
  \bibfield  {author} {\bibinfo {author} {\bibfnamefont {D.~P.}\ \bibnamefont
  {Feldman}},\ }\href@noop {} {\emph {\bibinfo {title} {Chaos and Fractals: An
  Elementary Introduction}}}\ (\bibinfo  {publisher} {Oxford University
  Press},\ \bibinfo {address} {Oxford},\ \bibinfo {year} {2012})\BibitemShut
  {NoStop}%
\bibitem [{\citenamefont {Palmer}\ \emph {et~al.}(2000)\citenamefont {Palmer},
  \citenamefont {Fairall},\ and\ \citenamefont {Brewer}}]{Palm00a}%
  \BibitemOpen
  \bibfield  {author} {\bibinfo {author} {\bibfnamefont {A.~J.}\ \bibnamefont
  {Palmer}}, \bibinfo {author} {\bibfnamefont {C.~W.}\ \bibnamefont {Fairall}},
  \ and\ \bibinfo {author} {\bibfnamefont {W.~A.}\ \bibnamefont {Brewer}},\
  }\href@noop {} {\bibfield  {journal} {\bibinfo  {journal} {IEEE Trans.\
  Geosci.\ Remote\ Sens.}\ }\textbf {\bibinfo {volume} {38}},\ \bibinfo {pages}
  {2056} (\bibinfo {year} {2000})}\BibitemShut {NoStop}%
\bibitem [{\citenamefont {Clarke}\ \emph {et~al.}(2003)\citenamefont {Clarke},
  \citenamefont {Freeman},\ and\ \citenamefont {Watkins}}]{Clar01a}%
  \BibitemOpen
  \bibfield  {author} {\bibinfo {author} {\bibfnamefont {R.~W.}\ \bibnamefont
  {Clarke}}, \bibinfo {author} {\bibfnamefont {M.~P.}\ \bibnamefont {Freeman}},
  \ and\ \bibinfo {author} {\bibfnamefont {N.~W.}\ \bibnamefont {Watkins}},\
  }\href@noop {} {\bibfield  {journal} {\bibinfo  {journal} {Phys. Rev. E}\
  }\textbf {\bibinfo {volume} {67}},\ \bibinfo {pages} {016203} (\bibinfo
  {year} {2003})}\BibitemShut {NoStop}%
\bibitem [{\citenamefont {Kelly}\ \emph {et~al.}(2012)\citenamefont {Kelly},
  \citenamefont {Dillingham}, \citenamefont {Hudson},\ and\ \citenamefont
  {Wiesner}}]{Kell12a}%
  \BibitemOpen
  \bibfield  {author} {\bibinfo {author} {\bibfnamefont {D.}~\bibnamefont
  {Kelly}}, \bibinfo {author} {\bibfnamefont {M.}~\bibnamefont {Dillingham}},
  \bibinfo {author} {\bibfnamefont {A.}~\bibnamefont {Hudson}}, \ and\ \bibinfo
  {author} {\bibfnamefont {K.}~\bibnamefont {Wiesner}},\ }\href@noop {}
  {\bibfield  {journal} {\bibinfo  {journal} {Public Library of Science One}\
  }\textbf {\bibinfo {volume} {7}},\ \bibinfo {pages} {e29703} (\bibinfo {year}
  {2012})}\BibitemShut {NoStop}%
\bibitem [{\citenamefont {Price}(1983)}]{Pric83a}%
  \BibitemOpen
  \bibfield  {author} {\bibinfo {author} {\bibfnamefont {G.~D.}\ \bibnamefont
  {Price}},\ }\href@noop {} {\bibfield  {journal} {\bibinfo  {journal} {Phys.\
  Chem.\ Minerals}\ }\textbf {\bibinfo {volume} {10}},\ \bibinfo {pages} {77}
  (\bibinfo {year} {1983})}\BibitemShut {NoStop}%
\bibitem [{\citenamefont {Ferraris}\ \emph {et~al.}(2008)\citenamefont
  {Ferraris}, \citenamefont {Makovicky},\ and\ \citenamefont
  {Merlino}}]{Ferr08a}%
  \BibitemOpen
  \bibfield  {author} {\bibinfo {author} {\bibfnamefont {G.}~\bibnamefont
  {Ferraris}}, \bibinfo {author} {\bibfnamefont {E.}~\bibnamefont {Makovicky}},
  \ and\ \bibinfo {author} {\bibfnamefont {S.}~\bibnamefont {Merlino}},\
  }\href@noop {} {\emph {\bibinfo {title} {Crystallography of Modular
  Materials}}},\ Vol.~\bibinfo {volume} {15}\ (\bibinfo  {publisher} {Oxford
  University Press},\ \bibinfo {year} {2008})\BibitemShut {NoStop}%
\bibitem [{\citenamefont {Varn}\ \emph {et~al.}(2002)\citenamefont {Varn},
  \citenamefont {Canright},\ and\ \citenamefont {Crutchfield}}]{Varn02a}%
  \BibitemOpen
  \bibfield  {author} {\bibinfo {author} {\bibfnamefont {D.~P.}\ \bibnamefont
  {Varn}}, \bibinfo {author} {\bibfnamefont {G.~S.}\ \bibnamefont {Canright}},
  \ and\ \bibinfo {author} {\bibfnamefont {J.~P.}\ \bibnamefont
  {Crutchfield}},\ }\href@noop {} {\bibfield  {journal} {\bibinfo  {journal}
  {Phys. Rev. B}\ }\textbf {\bibinfo {volume} {66}},\ \bibinfo {pages} {174110}
  (\bibinfo {year} {2002})}\BibitemShut {NoStop}%
\bibitem [{\citenamefont {Riechers}\ \emph {et~al.}(2014)\citenamefont
  {Riechers}, \citenamefont {Varn},\ and\ \citenamefont
  {Crutchfield}}]{Riec14b}%
  \BibitemOpen
  \bibfield  {author} {\bibinfo {author} {\bibfnamefont {P.~M.}\ \bibnamefont
  {Riechers}}, \bibinfo {author} {\bibfnamefont {D.~P.}\ \bibnamefont {Varn}},
  \ and\ \bibinfo {author} {\bibfnamefont {J.~P.}\ \bibnamefont
  {Crutchfield}},\ }\href@noop {} {\bibfield  {journal} {\bibinfo  {journal}
  {Santa Fe Institute Working Paper 2014-08-026}\ } (\bibinfo {year} {2014})},\
  \Eprint {http://arxiv.org/abs/1407.7159} {arXiv:1407.7159
  [cond-mat.mtrl-sci]} \BibitemShut {NoStop}%
\bibitem [{\citenamefont {Hendricks}\ and\ \citenamefont
  {Teller}(1942)}]{Hend42a}%
  \BibitemOpen
  \bibfield  {author} {\bibinfo {author} {\bibfnamefont {S.}~\bibnamefont
  {Hendricks}}\ and\ \bibinfo {author} {\bibfnamefont {E.}~\bibnamefont
  {Teller}},\ }\href@noop {} {\bibfield  {journal} {\bibinfo  {journal} {J.\
  Chem.\ Phys.}\ }\textbf {\bibinfo {volume} {10}},\ \bibinfo {pages} {147}
  (\bibinfo {year} {1942})}\BibitemShut {NoStop}%
\bibitem [{\citenamefont {Wilson}(1942)}]{Wils42a}%
  \BibitemOpen
  \bibfield  {author} {\bibinfo {author} {\bibfnamefont {A.~J.~C.}\
  \bibnamefont {Wilson}},\ }\href@noop {} {\bibfield  {journal} {\bibinfo
  {journal} {Proc.\ R.\ Soc.\ Ser.\ A}\ }\textbf {\bibinfo {volume} {180}},\
  \bibinfo {pages} {277} (\bibinfo {year} {1942})}\BibitemShut {NoStop}%
\bibitem [{\citenamefont {Kakinoki}\ and\ \citenamefont
  {Komura}(1965)}]{Kaki65a}%
  \BibitemOpen
  \bibfield  {author} {\bibinfo {author} {\bibfnamefont {J.}~\bibnamefont
  {Kakinoki}}\ and\ \bibinfo {author} {\bibfnamefont {Y.}~\bibnamefont
  {Komura}},\ }\href@noop {} {\bibfield  {journal} {\bibinfo  {journal} {Acta
  Crystallogr.}\ }\textbf {\bibinfo {volume} {19}},\ \bibinfo {pages} {137}
  (\bibinfo {year} {1965})}\BibitemShut {NoStop}%
\bibitem [{\citenamefont {Holloway}\ and\ \citenamefont
  {Klamkin}(1969)}]{Holl69a}%
  \BibitemOpen
  \bibfield  {author} {\bibinfo {author} {\bibfnamefont {H.}~\bibnamefont
  {Holloway}}\ and\ \bibinfo {author} {\bibfnamefont {M.~S.}\ \bibnamefont
  {Klamkin}},\ }\href@noop {} {\bibfield  {journal} {\bibinfo  {journal} {J.\
  Appl.\ Phys.}\ }\textbf {\bibinfo {volume} {40}},\ \bibinfo {pages} {1681}
  (\bibinfo {year} {1969})}\BibitemShut {NoStop}%
\bibitem [{\citenamefont {Holloway}(1969)}]{Holl69b}%
  \BibitemOpen
  \bibfield  {author} {\bibinfo {author} {\bibfnamefont {H.}~\bibnamefont
  {Holloway}},\ }\href@noop {} {\bibfield  {journal} {\bibinfo  {journal} {J.\
  Appl.\ Phys.}\ }\textbf {\bibinfo {volume} {40}},\ \bibinfo {pages} {4313}
  (\bibinfo {year} {1969})}\BibitemShut {NoStop}%
\bibitem [{\citenamefont {Pandey}\ \emph
  {et~al.}(1980{\natexlab{a}})\citenamefont {Pandey}, \citenamefont {Lele},\
  and\ \citenamefont {Krishna}}]{Pand80a}%
  \BibitemOpen
  \bibfield  {author} {\bibinfo {author} {\bibfnamefont {D.}~\bibnamefont
  {Pandey}}, \bibinfo {author} {\bibfnamefont {S.}~\bibnamefont {Lele}}, \ and\
  \bibinfo {author} {\bibfnamefont {P.}~\bibnamefont {Krishna}},\ }\href@noop
  {} {\bibfield  {journal} {\bibinfo  {journal} {Proc.\ R.\ Soc.\ London Ser.\
  A}\ }\textbf {\bibinfo {volume} {369}},\ \bibinfo {pages} {435} (\bibinfo
  {year} {1980}{\natexlab{a}})}\BibitemShut {NoStop}%
\bibitem [{\citenamefont {Pandey}\ \emph
  {et~al.}(1980{\natexlab{b}})\citenamefont {Pandey}, \citenamefont {Lele},\
  and\ \citenamefont {Krishna}}]{Pand80b}%
  \BibitemOpen
  \bibfield  {author} {\bibinfo {author} {\bibfnamefont {D.}~\bibnamefont
  {Pandey}}, \bibinfo {author} {\bibfnamefont {S.}~\bibnamefont {Lele}}, \ and\
  \bibinfo {author} {\bibfnamefont {P.}~\bibnamefont {Krishna}},\ }\href@noop
  {} {\bibfield  {journal} {\bibinfo  {journal} {Proc.\ R.\ Soc.\ London Ser.\
  A}\ }\textbf {\bibinfo {volume} {369}},\ \bibinfo {pages} {451} (\bibinfo
  {year} {1980}{\natexlab{b}})}\BibitemShut {NoStop}%
\bibitem [{\citenamefont {Pandey}\ \emph
  {et~al.}(1980{\natexlab{c}})\citenamefont {Pandey}, \citenamefont {Lele},\
  and\ \citenamefont {Krishna}}]{Pand80c}%
  \BibitemOpen
  \bibfield  {author} {\bibinfo {author} {\bibfnamefont {D.}~\bibnamefont
  {Pandey}}, \bibinfo {author} {\bibfnamefont {S.}~\bibnamefont {Lele}}, \ and\
  \bibinfo {author} {\bibfnamefont {P.}~\bibnamefont {Krishna}},\ }\href@noop
  {} {\bibfield  {journal} {\bibinfo  {journal} {Proc.\ R.\ Soc.\ London Ser.\
  A}\ }\textbf {\bibinfo {volume} {369}},\ \bibinfo {pages} {463} (\bibinfo
  {year} {1980}{\natexlab{c}})}\BibitemShut {NoStop}%
\bibitem [{\citenamefont {Sebastian}\ and\ \citenamefont
  {Krishna}(1994)}]{Seba94a}%
  \BibitemOpen
  \bibfield  {author} {\bibinfo {author} {\bibfnamefont {M.~T.}\ \bibnamefont
  {Sebastian}}\ and\ \bibinfo {author} {\bibfnamefont {P.}~\bibnamefont
  {Krishna}},\ }\href@noop {} {\emph {\bibinfo {title} {Random, Non-Random and
  Periodic Faulting in Crystals}}}\ (\bibinfo  {publisher} {Gordon and
  Breach},\ \bibinfo {address} {The Netherlands},\ \bibinfo {year}
  {1994})\BibitemShut {NoStop}%
\bibitem [{\citenamefont {Gosk}(2001)}]{Gosk01a}%
  \BibitemOpen
  \bibfield  {author} {\bibinfo {author} {\bibfnamefont {J.~B.}\ \bibnamefont
  {Gosk}},\ }\href@noop {} {\bibfield  {journal} {\bibinfo  {journal} {Crys.\
  Res.\ Tech.}\ }\textbf {\bibinfo {volume} {36}},\ \bibinfo {pages} {197}
  (\bibinfo {year} {2001})}\BibitemShut {NoStop}%
\bibitem [{\citenamefont {Kopp}\ \emph {et~al.}(2012)\citenamefont {Kopp},
  \citenamefont {Kaganer}, \citenamefont {Schwarzkopf}, \citenamefont
  {Waidick}, \citenamefont {Remmele}, \citenamefont {Kwasniewski},\ and\
  \citenamefont {Schmidbauer}}]{Kopp12a}%
  \BibitemOpen
  \bibfield  {author} {\bibinfo {author} {\bibfnamefont {V.~S.}\ \bibnamefont
  {Kopp}}, \bibinfo {author} {\bibfnamefont {V.~M.}\ \bibnamefont {Kaganer}},
  \bibinfo {author} {\bibfnamefont {J.}~\bibnamefont {Schwarzkopf}}, \bibinfo
  {author} {\bibfnamefont {F.}~\bibnamefont {Waidick}}, \bibinfo {author}
  {\bibfnamefont {T.}~\bibnamefont {Remmele}}, \bibinfo {author} {\bibfnamefont
  {A.}~\bibnamefont {Kwasniewski}}, \ and\ \bibinfo {author} {\bibfnamefont
  {M.}~\bibnamefont {Schmidbauer}},\ }\href@noop {} {\bibfield  {journal}
  {\bibinfo  {journal} {Acta Crystallogr.\ Sec.\ A}\ }\textbf {\bibinfo
  {volume} {68}},\ \bibinfo {pages} {148} (\bibinfo {year} {2012})}\BibitemShut
  {NoStop}%
\bibitem [{\citenamefont {Berliner}\ and\ \citenamefont
  {Werner}(1986)}]{Berl86a}%
  \BibitemOpen
  \bibfield  {author} {\bibinfo {author} {\bibfnamefont {R.}~\bibnamefont
  {Berliner}}\ and\ \bibinfo {author} {\bibfnamefont {S.}~\bibnamefont
  {Werner}},\ }\href@noop {} {\bibfield  {journal} {\bibinfo  {journal} {Phys.
  Rev. B}\ }\textbf {\bibinfo {volume} {34}},\ \bibinfo {pages} {3586}
  (\bibinfo {year} {1986})}\BibitemShut {NoStop}%
\bibitem [{\citenamefont {Kantz}\ and\ \citenamefont
  {Schreiber}(2004)}]{Kant04a}%
  \BibitemOpen
  \bibfield  {author} {\bibinfo {author} {\bibfnamefont {H.}~\bibnamefont
  {Kantz}}\ and\ \bibinfo {author} {\bibfnamefont {T.}~\bibnamefont
  {Schreiber}},\ }\href@noop {} {\emph {\bibinfo {title} {Nonlinear Time Series
  Analysis}}},\ \bibinfo {edition} {2nd}\ ed.\ (\bibinfo  {publisher}
  {Cambridge University Press},\ \bibinfo {address} {Cambridge},\ \bibinfo
  {year} {2004})\BibitemShut {NoStop}%
\bibitem [{\citenamefont {Nikolin}\ and\ \citenamefont
  {Babkevich}(1989)}]{Niko89a}%
  \BibitemOpen
  \bibfield  {author} {\bibinfo {author} {\bibfnamefont {B.~I.}\ \bibnamefont
  {Nikolin}}\ and\ \bibinfo {author} {\bibfnamefont {A.~Y.}\ \bibnamefont
  {Babkevich}},\ }\href@noop {} {\bibfield  {journal} {\bibinfo  {journal}
  {Acta Crystallogr.\ Sec.\ A}\ }\textbf {\bibinfo {volume} {45}},\ \bibinfo
  {pages} {797} (\bibinfo {year} {1989})}\BibitemShut {NoStop}%
\bibitem [{\citenamefont {Varn}(2001)}]{Varn01b}%
  \BibitemOpen
  \bibfield  {author} {\bibinfo {author} {\bibfnamefont {D.~P.}\ \bibnamefont
  {Varn}},\ }\emph {\bibinfo {title} {Language Extraction from ZnS}},\
  \href@noop {} {Ph.D. thesis},\ \bibinfo  {school} {University of Tennessee,
  Knoxville} (\bibinfo {year} {2001})\BibitemShut {NoStop}%
\bibitem [{\citenamefont {Varn}\ \emph
  {et~al.}(2013{\natexlab{a}})\citenamefont {Varn}, \citenamefont {Canright},\
  and\ \citenamefont {Crutchfield}}]{Varn13a}%
  \BibitemOpen
  \bibfield  {author} {\bibinfo {author} {\bibfnamefont {D.~P.}\ \bibnamefont
  {Varn}}, \bibinfo {author} {\bibfnamefont {G.~S.}\ \bibnamefont {Canright}},
  \ and\ \bibinfo {author} {\bibfnamefont {J.~P.}\ \bibnamefont
  {Crutchfield}},\ }\href@noop {} {\bibfield  {journal} {\bibinfo  {journal}
  {Acta Crystallogr.\ Sec.\ A}\ }\textbf {\bibinfo {volume} {69}},\ \bibinfo
  {pages} {197} (\bibinfo {year} {2013}{\natexlab{a}})}\BibitemShut {NoStop}%
\bibitem [{\citenamefont {Keen}\ and\ \citenamefont
  {McGreevy}(1990)}]{Keen90a}%
  \BibitemOpen
  \bibfield  {author} {\bibinfo {author} {\bibfnamefont {D.~A.}\ \bibnamefont
  {Keen}}\ and\ \bibinfo {author} {\bibfnamefont {R.~L.}\ \bibnamefont
  {McGreevy}},\ }\href@noop {} {\bibfield  {journal} {\bibinfo  {journal}
  {Nature}\ }\textbf {\bibinfo {volume} {344}},\ \bibinfo {pages} {423}
  (\bibinfo {year} {1990})}\BibitemShut {NoStop}%
\bibitem [{\citenamefont {Michels-Clark}\ \emph {et~al.}(2013)\citenamefont
  {Michels-Clark}, \citenamefont {Lynch}, \citenamefont {Hoffmann},
  \citenamefont {Hauser}, \citenamefont {Weber}, \citenamefont {Harrison},\
  and\ \citenamefont {B{\"{u}}rgi}}]{Mich13a}%
  \BibitemOpen
  \bibfield  {author} {\bibinfo {author} {\bibfnamefont {T.~M.}\ \bibnamefont
  {Michels-Clark}}, \bibinfo {author} {\bibfnamefont {V.~E.}\ \bibnamefont
  {Lynch}}, \bibinfo {author} {\bibfnamefont {C.~M.}\ \bibnamefont {Hoffmann}},
  \bibinfo {author} {\bibfnamefont {J.}~\bibnamefont {Hauser}}, \bibinfo
  {author} {\bibfnamefont {T.}~\bibnamefont {Weber}}, \bibinfo {author}
  {\bibfnamefont {R.}~\bibnamefont {Harrison}}, \ and\ \bibinfo {author}
  {\bibfnamefont {H.~B.}\ \bibnamefont {B{\"{u}}rgi}},\ }\href@noop {}
  {\bibfield  {journal} {\bibinfo  {journal} {J.\ Appl.\ Crystallogr.}\
  }\textbf {\bibinfo {volume} {46}},\ \bibinfo {pages} {1616} (\bibinfo {year}
  {2013})}\BibitemShut {NoStop}%
\bibitem [{\citenamefont {Johnson}\ \emph {et~al.}(2010)\citenamefont
  {Johnson}, \citenamefont {Crutchield}, \citenamefont {Ellison},\ and\
  \citenamefont {McTague}}]{John10b}%
  \BibitemOpen
  \bibfield  {author} {\bibinfo {author} {\bibfnamefont {B.~D.}\ \bibnamefont
  {Johnson}}, \bibinfo {author} {\bibfnamefont {J.~P.}\ \bibnamefont
  {Crutchield}}, \bibinfo {author} {\bibfnamefont {C.~J.}\ \bibnamefont
  {Ellison}}, \ and\ \bibinfo {author} {\bibfnamefont {C.~S.}\ \bibnamefont
  {McTague}},\ }\href@noop {} {\bibfield  {journal} {\bibinfo  {journal} {Santa
  Fe Institute Working Paper 10-11-027}\ } (\bibinfo {year} {2010})},\ \Eprint
  {http://arxiv.org/abs/1011.0036v3} {arXiv:1011.0036v3 [cs.FL]} \BibitemShut
  {NoStop}%
\bibitem [{\citenamefont {Strelioff}\ and\ \citenamefont
  {Crutchfield}(2014)}]{Stre14a}%
  \BibitemOpen
  \bibfield  {author} {\bibinfo {author} {\bibfnamefont {C.~C.}\ \bibnamefont
  {Strelioff}}\ and\ \bibinfo {author} {\bibfnamefont {J.~P.}\ \bibnamefont
  {Crutchfield}},\ }\href@noop {} {\bibfield  {journal} {\bibinfo  {journal}
  {Phys. Rev. E}\ }\textbf {\bibinfo {volume} {89}},\ \bibinfo {pages} {042119}
  (\bibinfo {year} {2014})}\BibitemShut {NoStop}%
\bibitem [{\citenamefont {Badii}\ and\ \citenamefont {Politi}(1997)}]{Badi97a}%
  \BibitemOpen
  \bibfield  {author} {\bibinfo {author} {\bibfnamefont {R.}~\bibnamefont
  {Badii}}\ and\ \bibinfo {author} {\bibfnamefont {A.}~\bibnamefont {Politi}},\
  }\href@noop {} {\emph {\bibinfo {title} {Complexity: Hierarchical Structures
  and Scaling in Physics}}},\ \bibinfo {series} {Cambridge Nonlinear Science
  Series}, Vol.~\bibinfo {volume} {6}\ (\bibinfo  {publisher} {Cambridge
  University Press},\ \bibinfo {year} {1997})\BibitemShut {NoStop}%
\bibitem [{\citenamefont {Varn}\ and\ \citenamefont
  {Crutchfield}(2004)}]{Varn04a}%
  \BibitemOpen
  \bibfield  {author} {\bibinfo {author} {\bibfnamefont {D.~P.}\ \bibnamefont
  {Varn}}\ and\ \bibinfo {author} {\bibfnamefont {J.~P.}\ \bibnamefont
  {Crutchfield}},\ }\href@noop {} {\bibfield  {journal} {\bibinfo  {journal}
  {Phys.\ Lett.\ A}\ }\textbf {\bibinfo {volume} {324}},\ \bibinfo {pages}
  {299} (\bibinfo {year} {2004})}\BibitemShut {NoStop}%
\bibitem [{\citenamefont {Baake}\ and\ \citenamefont {Grimm}(2012)}]{Baak12a}%
  \BibitemOpen
  \bibfield  {author} {\bibinfo {author} {\bibfnamefont {M.}~\bibnamefont
  {Baake}}\ and\ \bibinfo {author} {\bibfnamefont {U.}~\bibnamefont {Grimm}},\
  }\href@noop {} {\bibfield  {journal} {\bibinfo  {journal} {Chem.\ Soc.\ Rev}\
  }\textbf {\bibinfo {volume} {41}},\ \bibinfo {pages} {6821–6843} (\bibinfo
  {year} {2012})}\BibitemShut {NoStop}%
\bibitem [{\citenamefont {Ashcroft}\ and\ \citenamefont
  {Mermin}(1976)}]{Ashc76a}%
  \BibitemOpen
  \bibfield  {author} {\bibinfo {author} {\bibfnamefont {N.~W.}\ \bibnamefont
  {Ashcroft}}\ and\ \bibinfo {author} {\bibfnamefont {N.~D.}\ \bibnamefont
  {Mermin}},\ }\href@noop {} {\emph {\bibinfo {title} {Solid State Physics}}}\
  (\bibinfo  {publisher} {Saunders College Publishing},\ \bibinfo {address}
  {New York},\ \bibinfo {year} {1976})\BibitemShut {NoStop}%
\bibitem [{\citenamefont {Ortiz}\ \emph {et~al.}(2013)\citenamefont {Ortiz},
  \citenamefont {S{\'{a}}nchez-Bajo}, \citenamefont {Cumbrera},\ and\
  \citenamefont {Guiberteau}}]{Orti13a}%
  \BibitemOpen
  \bibfield  {author} {\bibinfo {author} {\bibfnamefont {A.~L.}\ \bibnamefont
  {Ortiz}}, \bibinfo {author} {\bibfnamefont {F.}~\bibnamefont
  {S{\'{a}}nchez-Bajo}}, \bibinfo {author} {\bibfnamefont {F.~L.}\ \bibnamefont
  {Cumbrera}}, \ and\ \bibinfo {author} {\bibfnamefont {F.}~\bibnamefont
  {Guiberteau}},\ }\href@noop {} {\bibfield  {journal} {\bibinfo  {journal}
  {J.\ Appl.\ Crystallogr.}\ }\textbf {\bibinfo {volume} {46}},\ \bibinfo
  {pages} {242} (\bibinfo {year} {2013})}\BibitemShut {NoStop}%
\bibitem [{Note1()}]{Note1}%
  \BibitemOpen
  \bibinfo {note} {For power spectra of finite-length sequences, there is no
  clear distinction among these features.}\BibitemShut {Stop}%
\bibitem [{\citenamefont {Axel}\ and\ \citenamefont
  {Terauchi}(1991)}]{Axel91a}%
  \BibitemOpen
  \bibfield  {author} {\bibinfo {author} {\bibfnamefont {F.}~\bibnamefont
  {Axel}}\ and\ \bibinfo {author} {\bibfnamefont {H.}~\bibnamefont
  {Terauchi}},\ }\href@noop {} {\bibfield  {journal} {\bibinfo  {journal}
  {Phys. Rev. Lett.}\ }\textbf {\bibinfo {volume} {66}},\ \bibinfo {pages}
  {2223} (\bibinfo {year} {1991})}\BibitemShut {NoStop}%
\bibitem [{\citenamefont {Kabra}\ and\ \citenamefont {Pandey}(1988)}]{Kabr88a}%
  \BibitemOpen
  \bibfield  {author} {\bibinfo {author} {\bibfnamefont {V.~K.}\ \bibnamefont
  {Kabra}}\ and\ \bibinfo {author} {\bibfnamefont {D.}~\bibnamefont {Pandey}},\
  }\href@noop {} {\bibfield  {journal} {\bibinfo  {journal} {Phys. Rev. Lett.}\
  }\textbf {\bibinfo {volume} {61}},\ \bibinfo {pages} {1493} (\bibinfo {year}
  {1988})}\BibitemShut {NoStop}%
\bibitem [{\citenamefont {Yi}\ and\ \citenamefont {Canright}(1996)}]{Yi96a}%
  \BibitemOpen
  \bibfield  {author} {\bibinfo {author} {\bibfnamefont {J.}~\bibnamefont
  {Yi}}\ and\ \bibinfo {author} {\bibfnamefont {G.~S.}\ \bibnamefont
  {Canright}},\ }\href@noop {} {\bibfield  {journal} {\bibinfo  {journal}
  {Phys. Rev. B}\ }\textbf {\bibinfo {volume} {53}},\ \bibinfo {pages} {5198}
  (\bibinfo {year} {1996})}\BibitemShut {NoStop}%
\bibitem [{\citenamefont {Varn}\ and\ \citenamefont
  {Canright}(2001)}]{Varn01a}%
  \BibitemOpen
  \bibfield  {author} {\bibinfo {author} {\bibfnamefont {D.~P.}\ \bibnamefont
  {Varn}}\ and\ \bibinfo {author} {\bibfnamefont {G.~S.}\ \bibnamefont
  {Canright}},\ }\href@noop {} {\bibfield  {journal} {\bibinfo  {journal} {Acta
  Crystallogr.\ Sec.\ A}\ }\textbf {\bibinfo {volume} {57}},\ \bibinfo {pages}
  {4} (\bibinfo {year} {2001})}\BibitemShut {NoStop}%
\bibitem [{Note2()}]{Note2}%
  \BibitemOpen
  \bibinfo {note} {As previously done,~\cite {Varn13a} we divide out those
  factors associated with experimental corrections to the observed DP, as well
  as the total number ${N}$ of MLs, so that ${\protect \sf {I}}(\ell )$ has
  only those contributions arising from the stacking structure itself. Here and
  elsewhere, we refer to ${\protect \sf {I}}(\ell )$ simply as the
  DP.}\BibitemShut {Stop}%
\bibitem [{\citenamefont {Estevez-Rams}\ \emph {et~al.}(2003)\citenamefont
  {Estevez-Rams}, \citenamefont {Aragon-Fernandez}, \citenamefont {Fuess},\
  and\ \citenamefont {Penton-Madrigal}}]{Este03a}%
  \BibitemOpen
  \bibfield  {author} {\bibinfo {author} {\bibfnamefont {E.}~\bibnamefont
  {Estevez-Rams}}, \bibinfo {author} {\bibfnamefont {B.}~\bibnamefont
  {Aragon-Fernandez}}, \bibinfo {author} {\bibfnamefont {H.}~\bibnamefont
  {Fuess}}, \ and\ \bibinfo {author} {\bibfnamefont {A.}~\bibnamefont
  {Penton-Madrigal}},\ }\href@noop {} {\bibfield  {journal} {\bibinfo
  {journal} {Phys. Rev. B}\ }\textbf {\bibinfo {volume} {68}},\ \bibinfo
  {pages} {064111} (\bibinfo {year} {2003})}\BibitemShut {NoStop}%
\bibitem [{Note3()}]{Note3}%
  \BibitemOpen
  \bibinfo {note} {It may seem that specializing to such a specific expression
  for the DP at this stage limits the applicability of the approach. While the
  development here is restricted to the case of CPS, under mild conditions, the
  Wiener-Khinchin theorem~\cite {Badi97a} guarantees that power spectra can be
  written in terms of pair autocorrelation functions, as is done here. This
  makes the spectral decomposition rather generic.}\BibitemShut {Stop}%
\bibitem [{\citenamefont {Riechers}\ and\ \citenamefont
  {Crutchfield}(2014)}]{Riec14a}%
  \BibitemOpen
  \bibfield  {author} {\bibinfo {author} {\bibfnamefont {P.~M.}\ \bibnamefont
  {Riechers}}\ and\ \bibinfo {author} {\bibfnamefont {J.~P.}\ \bibnamefont
  {Crutchfield}},\ }\href@noop {} {\enquote {\bibinfo {title} {Spectral
  decomposition of structural complexity:~{M}eromorphic functional calculus of
  nondiagonalizable dynamics},}\ } (\bibinfo {year} {2014}),\ \bibinfo {note}
  {manuscript in preparation}\BibitemShut {NoStop}%
\bibitem [{\citenamefont {Oppenheim}\ and\ \citenamefont
  {Schafer}(1975)}]{Oppe75a}%
  \BibitemOpen
  \bibfield  {author} {\bibinfo {author} {\bibfnamefont {A.~V.}\ \bibnamefont
  {Oppenheim}}\ and\ \bibinfo {author} {\bibfnamefont {R.~W.}\ \bibnamefont
  {Schafer}},\ }\href@noop {} {\emph {\bibinfo {title} {Digital Signal
  Processing}}}\ (\bibinfo  {publisher} {Prentice-Hall},\ \bibinfo {address}
  {Englewood Cliffs},\ \bibinfo {year} {1975})\BibitemShut {NoStop}%
\bibitem [{Note4()}]{Note4}%
  \BibitemOpen
  \bibinfo {note} {By magnitude, we mean the $\ell $-integral over the $\delta
  $-function. If integrating with respect to a related variable, then the
  magnitude of the $\delta $-function changes accordingly. As a simple example,
  integrating over $\omega = 2\pi \ell $ changes the magnitude of the $\delta
  $-function by a factor of $2 \pi $.}\BibitemShut {Stop}%
\bibitem [{Note5()}]{Note5}%
  \BibitemOpen
  \bibinfo {note} {$\protect \Braket {{\protect \mathcal {T}}_{\lambda , m}
  ^{{\protect \mathaccentV {hat}05E{\xi }}({\protect \mathcal A})}}$ is
  constant with respect to the relative layer displacement $n$. However,
  ${\setbox \z@ \hbox {\frozen@everymath \@emptytoks \mathsurround \z@
  $\nulldelimiterspace \z@ \left \protect \{\vcenter to1.5\big@size {}\right
  .$}\box \z@ } \protect \Braket {{\protect \mathcal {T}}_{\lambda , m}
  ^{{\protect \mathaccentV {hat}05E{\xi }}({\protect \mathcal A})}}{\setbox \z@
  \hbox {\frozen@everymath \@emptytoks \mathsurround \z@ $\nulldelimiterspace
  \z@ \left \protect \}\vcenter to1.5\big@size {}\right .$}\box \z@ }$ can be a
  function of a process's parameters.}\BibitemShut {Stop}%
\bibitem [{Note6()}]{Note6}%
  \BibitemOpen
  \bibinfo {note} {Here and in the following examples, we define a bar over a
  variable to mean one minus that variable: $\protect \mathaccentV {bar}016{x}
  \equiv 1-x$.}\BibitemShut {Stop}%
\bibitem [{\citenamefont {Warren}(1969)}]{Warr69a}%
  \BibitemOpen
  \bibfield  {author} {\bibinfo {author} {\bibfnamefont {B.~E.}\ \bibnamefont
  {Warren}},\ }\href@noop {} {\emph {\bibinfo {title} {X-Ray Diffraction}}}\
  (\bibinfo  {publisher} {Addison-Wesley},\ \bibinfo {year} {1969})\BibitemShut
  {NoStop}%
\bibitem [{\citenamefont {Guinier}(1963)}]{Guin63a}%
  \BibitemOpen
  \bibfield  {author} {\bibinfo {author} {\bibfnamefont {A.}~\bibnamefont
  {Guinier}},\ }\href@noop {} {\emph {\bibinfo {title} {X-Ray Diffraction in
  Crystals, Imperfect Crystals, and Amorphous Bodies}}}\ (\bibinfo  {publisher}
  {W.~H. Freeman and Company},\ \bibinfo {address} {New York},\ \bibinfo {year}
  {1963})\BibitemShut {NoStop}%
\bibitem [{\citenamefont {Estevez-Rams}\ \emph {et~al.}(2008)\citenamefont
  {Estevez-Rams}, \citenamefont {Welzel}, \citenamefont {Madrigal},\ and\
  \citenamefont {Mittemeijer}}]{Este08a}%
  \BibitemOpen
  \bibfield  {author} {\bibinfo {author} {\bibfnamefont {E.}~\bibnamefont
  {Estevez-Rams}}, \bibinfo {author} {\bibfnamefont {U.}~\bibnamefont
  {Welzel}}, \bibinfo {author} {\bibfnamefont {A.~P.}\ \bibnamefont
  {Madrigal}}, \ and\ \bibinfo {author} {\bibfnamefont {E.~J.}\ \bibnamefont
  {Mittemeijer}},\ }\href@noop {} {\bibfield  {journal} {\bibinfo  {journal}
  {Acta Crystallogr.\ Sec.\ A}\ }\textbf {\bibinfo {volume} {64}},\ \bibinfo
  {pages} {537} (\bibinfo {year} {2008})}\BibitemShut {NoStop}%
\bibitem [{\citenamefont {Sun}\ \emph {et~al.}(2012)\citenamefont {Sun},
  \citenamefont {Robert}, \citenamefont {Andreadou}, \citenamefont {Mantzari},
  \citenamefont {Jokubavicius}, \citenamefont {Yakimova}, \citenamefont
  {Camassel}, \citenamefont {Juillaguet}, \citenamefont {Polychroniadis},\ and\
  \citenamefont {Syv{\"{a}}j{\"{a}}rvi}}]{Sun12a}%
  \BibitemOpen
  \bibfield  {author} {\bibinfo {author} {\bibfnamefont {J.~W.}\ \bibnamefont
  {Sun}}, \bibinfo {author} {\bibfnamefont {T.}~\bibnamefont {Robert}},
  \bibinfo {author} {\bibfnamefont {A.}~\bibnamefont {Andreadou}}, \bibinfo
  {author} {\bibfnamefont {A.}~\bibnamefont {Mantzari}}, \bibinfo {author}
  {\bibfnamefont {V.}~\bibnamefont {Jokubavicius}}, \bibinfo {author}
  {\bibfnamefont {R.}~\bibnamefont {Yakimova}}, \bibinfo {author}
  {\bibfnamefont {J.}~\bibnamefont {Camassel}}, \bibinfo {author}
  {\bibfnamefont {S.}~\bibnamefont {Juillaguet}}, \bibinfo {author}
  {\bibfnamefont {E.~K.}\ \bibnamefont {Polychroniadis}}, \ and\ \bibinfo
  {author} {\bibfnamefont {M.}~\bibnamefont {Syv{\"{a}}j{\"{a}}rvi}},\
  }\href@noop {} {\bibfield  {journal} {\bibinfo  {journal} {J.\ Appl.\ Phys.}\
  }\textbf {\bibinfo {volume} {111}},\ \bibinfo {pages} {113527} (\bibinfo
  {year} {2012})}\BibitemShut {NoStop}%
\bibitem [{\citenamefont {Crutchfield}\ and\ \citenamefont
  {Young}(1989)}]{Crut89a}%
  \BibitemOpen
  \bibfield  {author} {\bibinfo {author} {\bibfnamefont {J.~P.}\ \bibnamefont
  {Crutchfield}}\ and\ \bibinfo {author} {\bibfnamefont {K.}~\bibnamefont
  {Young}},\ }\href@noop {} {\bibfield  {journal} {\bibinfo  {journal} {Phys.
  Rev. Lett.}\ }\textbf {\bibinfo {volume} {63}},\ \bibinfo {pages} {105}
  (\bibinfo {year} {1989})}\BibitemShut {NoStop}%
\bibitem [{\citenamefont {Varn}\ \emph {et~al.}(2007)\citenamefont {Varn},
  \citenamefont {Canright},\ and\ \citenamefont {Crutchfield}}]{Varn07a}%
  \BibitemOpen
  \bibfield  {author} {\bibinfo {author} {\bibfnamefont {D.~P.}\ \bibnamefont
  {Varn}}, \bibinfo {author} {\bibfnamefont {G.~S.}\ \bibnamefont {Canright}},
  \ and\ \bibinfo {author} {\bibfnamefont {J.~P.}\ \bibnamefont
  {Crutchfield}},\ }\href@noop {} {\bibfield  {journal} {\bibinfo  {journal}
  {Acta Crystallogr.\ Sec.\ B}\ }\textbf {\bibinfo {volume} {63}},\ \bibinfo
  {pages} {169} (\bibinfo {year} {2007})}\BibitemShut {NoStop}%
\bibitem [{\citenamefont {Varn}\ \emph
  {et~al.}(2013{\natexlab{b}})\citenamefont {Varn}, \citenamefont {Canright},\
  and\ \citenamefont {Crutchfield}}]{Varn13b}%
  \BibitemOpen
  \bibfield  {author} {\bibinfo {author} {\bibfnamefont {D.~P.}\ \bibnamefont
  {Varn}}, \bibinfo {author} {\bibfnamefont {G.~S.}\ \bibnamefont {Canright}},
  \ and\ \bibinfo {author} {\bibfnamefont {J.~P.}\ \bibnamefont
  {Crutchfield}},\ }\href@noop {} {\bibfield  {journal} {\bibinfo  {journal}
  {Acta Crystallogr.\ Sec.\ A}\ }\textbf {\bibinfo {volume} {69}},\ \bibinfo
  {pages} {413} (\bibinfo {year} {2013}{\natexlab{b}})}\BibitemShut {NoStop}%
\bibitem [{\citenamefont {Crutchfield}\ \emph {et~al.}(2013)\citenamefont
  {Crutchfield}, \citenamefont {Ellison},\ and\ \citenamefont
  {Riechers}}]{Crut13a}%
  \BibitemOpen
  \bibfield  {author} {\bibinfo {author} {\bibfnamefont {J.~P.}\ \bibnamefont
  {Crutchfield}}, \bibinfo {author} {\bibfnamefont {C.~J.}\ \bibnamefont
  {Ellison}}, \ and\ \bibinfo {author} {\bibfnamefont {P.~M.}\ \bibnamefont
  {Riechers}},\ }\href@noop {} {\bibfield  {journal} {\bibinfo  {journal}
  {Santa Fe Institute Working Paper 2013-09-028}\ } (\bibinfo {year} {2013})},\
  \Eprint {http://arxiv.org/abs/1309.3792} {arXiv:1309.3792
  [cond-mat.stat-mech]} \BibitemShut {NoStop}%
\end{thebibliography}%

\end{document}